\documentclass[sigconf]{acmart}

\AtBeginDocument{%
  }

\copyrightyear{2023}
\acmYear{2023}
\setcopyright{rightsretained}
\acmConference[WWW '23]{Proceedings of the ACM Web Conference 2023}{April 30-May 4, 2023}{Austin, TX, USA}
\acmBooktitle{Proceedings of the ACM Web Conference 2023 (WWW '23), April 30-May 4, 2023, Austin, TX, USA}
\acmDOI{10.1145/3543507.3583251}
\acmISBN{978-1-4503-9416-1/23/04}

\usepackage{algorithmic}
\usepackage{graphicx}
\usepackage{xspace}
\usepackage{multirow}
\usepackage[caption=false]{subfig}

\newcommand{\bm}{\texttt{BM3}}

\newcommand{\ie}{\emph{i.e.},\xspace}
\newcommand{\eg}{\emph{e.g.},\xspace}

\begin{document}

\title{Bootstrap Latent Representations for Multi-modal Recommendation}

\author{Xin Zhou}
\email{xin.zhou@ntu.edu.sg}
\affiliation{%
	\institution{Nanyang Technological University}
	\city{Singapore}
	\country{Singapore}
}

\author{Hongyu Zhou}
\email{hongyu.zhou@ntu.edu.sg}
\affiliation{%
	\institution{Nanyang Technological University}
	\city{Singapore}
	\country{Singapore}
}

\author{Yong Liu}
\email{stephenliu@ntu.edu.sg}
\affiliation{%
	\institution{Nanyang Technological University}
	\city{Singapore}
	\country{Singapore}
}

\author{Zhiwei Zeng}
\email{zhiwei.zeng@ntu.edu.sg}
\affiliation{%
	\institution{Nanyang Technological University}
	\city{Singapore}
	\country{Singapore}
}

\author{Chunyan Miao}
\email{ascymiao@ntu.edu.sg}
\affiliation{%
	\institution{Nanyang Technological University}
	\city{Singapore}
	\country{Singapore}
}

\author{Pengwei Wang}
\email{hoverwang.wpw@alibaba-inc.com}
\affiliation{%
	\institution{Alibaba Group}
	\city{Beijing}
	\country{China}
}

\author{Yuan You}
\email{youyuan.yy@alibaba-inc.com}
\affiliation{%
	\institution{Alibaba Group}
	\city{Hangzhou}
	\country{China}
}

\author{Feijun Jiang}
\email{feijun.jiangfj@antgroup.com}
\affiliation{%
	\institution{Alibaba Group}
	\city{Hangzhou}
	\country{China}
}

\renewcommand{\shortauthors}{Xin Zhou, et al.}

\begin{abstract}
This paper studies the multi-modal recommendation problem, where the item multi-modality information (\eg images and textual descriptions) is exploited to improve the recommendation accuracy. Besides the user-item interaction graph, existing state-of-the-art methods usually use auxiliary graphs (\eg user-user or item-item relation graph) to augment the learned representations of users and/or items.
These representations are often propagated and aggregated on auxiliary graphs using graph convolutional networks, which can be prohibitively expensive in computation and memory, especially for large graphs.
Moreover, existing multi-modal recommendation methods usually leverage randomly sampled negative examples in Bayesian Personalized Ranking (BPR) loss to guide the learning of user/item representations, which increases the computational cost on large graphs and may also bring noisy supervision signals into the training process.
To tackle the above issues, we propose a novel self-supervised multi-modal recommendation model, 
dubbed \bm{}, which requires neither augmentations from auxiliary graphs nor negative samples.
Specifically, \bm{} first bootstraps latent contrastive views from the representations of users and items with a simple dropout augmentation. 
It then jointly optimizes three multi-modal objectives to learn the representations of users and items by reconstructing the user-item interaction graph and aligning modality features under both inter- and intra-modality perspectives. 
\bm{} alleviates both the need for contrasting with negative examples and the complex graph augmentation from an additional target network for contrastive view generation. 
We show \bm{} outperforms prior recommendation models on three datasets with number of nodes ranging from 20K to 200K, while achieving a 2-9$\times$ reduction in training time. Code implementation is located at: \url{https://github.com/enoche/BM3}.
\end{abstract}


\maketitle

\section{Introduction}

In fast-growing e-commerce businesses, recommender systems play a critical role in helping users discover products or services they may like among millions of offerings. In practice, deep learning techniques have been widely applied in recommendation systems, mainly for exploiting historical user-item interactions, to model users' preferences on items and produce item recommendations to users~\cite{zhang2019deep}. However, the rich multi-modal content information (\eg texts, images, and videos) of items has still not been fully explored. 

To improve the recommendation accuracy, recent work on multi-modal recommendation have studied effective means to integrate item multi-modal information into the traditional user-item recommendation paradigm.
For example, some methods concatenate multi-modal features with the latent representations of items~\cite{he2016vbpr} or leverage attention mechanisms~\cite{liu2019user, chen2019personalized} to capture users' preferences on items' multi-modal features. With a surge of research on graph-based recommendations~\cite{wu2020graph,wang2021graph,zhou2022layer}, another line of research uses Graph Neural Networks (GNNs) to exploit item multi-modal information and enhance the learning of user and item representations~\cite{wei2019mmgcn,wei2020graph,wang2021dualgnn}. For instance, \cite{wei2019mmgcn} uses graph convolutional networks to separately propagate and aggregate different item multi-modal information on the user-item interaction graph. To further improve recommendation performance, other auxiliary graph structures, \eg the user-user relation graph~\cite{wang2021dualgnn} and item-item relation graph~\cite{zhang2021mining}, have also been exploited to enhance the learning of user and item representations from the multi-modal information. 

Although existing GNN-based multi-modal methods~\cite{wei2020graph,wang2021dualgnn,zhang2021mining,zhou2022tale,zhou2023enhancing} can achieve state-of-the-art recommendation accuracy, the following issues may hinder their applications in scenarios involving large-scale graphs.
\textit{First}, they often learn the user and item representations based on pair-wise ranking losses, \eg the Bayesian Personalized Ranking (BPR) loss~\cite{rendle2009bpr}, which treat observed user-item interaction pairs as positive samples and randomly sampled user-item pairs as negative samples. Such a negative sampling strategy may incur a prohibitive cost on large graphs~\cite{thakoor2021large} and bring noisy supervision signals into the training process. For example, previous research~\cite{zhou2021selfcf} has verified that the default uniform sampling~\cite{zhang2013optimizing} in LightGCN~\cite{he2020lightgcn} obsesses more than 25\% training time per epoch.
\textit{Second}, methods utilizing auxiliary graph structures may incur prohibitive memory cost when building and/or training on large-scale auxiliary graphs. More analyses on the computational complexity of existing graph-based multi-modal methods can be found in Table~\ref{tab:com_com} and Table~\ref{tab:perform_eff}.

Self-Supervised Learning (SSL)~\cite{grill2020bootstrap, chen2021exploring} provides a possible solution for learning the representations of users and items without negative samples. 
Research in various domains ranging from Computer Vision (CV) to Natural Language Processing (NLP), has shown that SSL is possible to achieve competitive or even better results than supervised learning~\cite{grill2020bootstrap, chent2020simple, zbontar2021barlow}. 
The main idea of SSL is to maximize the similarity of representations obtained from different \textit{distorted versions} of a sample using two asymmetry networks, \ie the online network and the target network.
However, training with only positive samples will lead the model into a trivial constant solution~\cite{chen2021exploring}. 
To tackle this collapsing problem, BYOL~\cite{grill2020bootstrap} and SimSiam~\cite{chen2021exploring} introduce an additional ``predictor'' network to the online network and a special ``stop gradient'' operation on the target network.
Recently, BUIR~\cite{lee2021bootstrapping} transfers BYOL into the recommendation domain and shows competitive performance on the evaluation datasets.

In this paper, we propose a \textbf{\underline{B}}ootstrapped \textbf{\underline{M}}ulti-\textbf{\underline{M}}odal \textbf{\underline{M}}odel, dubbed \bm{}, for multi-modal recommendation. 
It first simplifies the current SSL framework by removing its target network, which can reduce half of the model parameters.
Moreover, to retain the similarity between different augmentations, \bm{} incorporates a simple dropout mechanism to perturb the latent embeddings generated from the online network. This is different from current SSL paradigm that perturbs the inputs via graph augmentation~\cite{grill2020bootstrap, lee2021bootstrapping} or image augmentation~\cite{shorten2019survey}.
The design eases both the memory and computational cost of conventional graph augmentation techniques~\cite{lee2021bootstrapping, wu2021self}, as it does not introduce any auxiliary graphs.
Last but not least, we design a loss function that is specialized for multi-modal recommendation. It minimizes the reconstruction loss of the user-item interaction graph as well as aligns the learned features under both inter- and intra-modality perspectives.

We summarize our main contributions as follows. \textit{First}, we propose \bm{}, a novel self-supervised learning method for multi-modal recommendation. 
	In \bm{}, we use a simple latent representation dropout mechanism instead of graph augmentation to generate the target view of a user or an item for contrastive learning without negative samples. 
\textit{Second}, to train \bm{} without negative samples, we design a Multi-Modal Contrastive Loss (MMCL) function that jointly optimizes three objectives. In addition to minimizing the classic user-item interaction graph reconstruction loss, MMCL further aligns the learned features between different modalities and reduces the dissimilarity between representations of different augmented views from a specific modality.
\textit{Finally}, we validate the effectiveness and efficiency of \bm{} on three datasets with the number of nodes ranging from 20K to 200K. The experimental results show that \bm{} achieves significant improvements over the state-of-the-art multi-modal recommendation methods, while training 2-9$\times$ faster than the baseline methods.

\section{Related Work}
\label{sec:relatedwork}
\subsection{Multi-modal Recommendation}
\subsubsection{Deep Learning-based Models}
Due to the success of the Collaborative Filtering (CF) method, most early multi-modal recommendation models utilize deep learning techniques to explore users' preferences on top of the CF paradigm.
For example, VBPR~\cite{he2016vbpr}, which builds on top of the BPR method, leverages the visual features of items. It utilizes a pre-trained convolutional neural network to obtain the visual features of items and linearly transforms them into a latent visual space. To make predictions, VBPR represents an item by concatenating the latent visual features with its ID embedding. 
Moreover, Deepstyle~\cite{liu2017deepstyle} augments the representations of items with both visual and style features within the BPR framework.
To capture the users' preference on multi-modal information, the attention mechanism has also been adopted in recommendation models.
For instance, VECF~\cite{chen2019personalized} utilizes the VGG model~\cite{simonyan2014very} to perform pre-segmentation on images and captures the user's attention on different image regions.
MAML~\cite{liu2019user} uses a two-layer neural network to capture the user's preference on textual and visual features of an item.

\subsubsection{Graph-based Multi-modal Models}

More recently, another line of research introduces GNNs into recommendation systems~\cite{wu2020graph}, which can greatly enhance the user and item representations by incorporating the structural information in the user-item interaction graph and auxiliary graphs.

To exploit the item multi-modal information, MMGCN~\cite{wei2019mmgcn} adopts the message passing mechanism of Graph Convolutional Networks (GCNs) and constructs a modality-specific user-item bipartite graph, which can capture the information from multi-hop neighbors to enhance the user and item representations.
Based on MMGCN, DualGNN~\cite{wang2021dualgnn} introduces a user co-occurrence graph with a model preference learning module to capture the user's preference for features from different modalities of an item. 
As the user-item graph may encompass unintentional interactions, GRCN~\cite{wei2020graph} introduces a graph refine layer to refine the structure of the user-item interaction graph by identifying the noise edges and corrupting the false-positive edges. 
To explicitly mine the semantic information between items, LATTICE~\cite{zhang2021mining} constructs item-item relation graphs for each modality and fuses them together to obtain a latent item graph. It dynamically updates the graph after items' information is propagated and aggregated from their highly connected affinities using GCNs.
FREEDOM~\cite{zhou2022tale} further detects that the learning of item-item graphs are negligible and freezes the graph for effective and efficient recommendation.
\cite{zhou2023comprehensive} provides a comprehensive survey of multi-modal recommender systems with taxonomy, evaluation and future directions.

Although graph-based multi-modal models achieve new state-of-the-art recommendation accuracy, they often require auxiliary graphs for user and item augmentations and also a large number of negatives for representation learning with BPR loss.
Both requirements can lead to high computational complexity and prohibitive memory cost as the graph size increases, limiting the efficiency of these models in scenarios involving large-scale graphs.

\subsection{Self-supervised Learning (SSL)}
SSL-based methods have achieved competitive results in various CV and NLP tasks~\cite{jing2020self,liu2021self}.
As our model is based on SSL that only uses observed data, our review of SSL methods focuses on those that do not require negative sampling.

Current SSL frameworks are derived from Siamese networks~\cite{bromley1993signature}, which are generic models for comparing entities~\cite{chen2021exploring}.
BYOL~\cite{grill2020bootstrap} and SimSiam~\cite{chen2021exploring} use asymmetric Siamese network to achieve remarkable results.
Specifically, BYOL proposes two coupled encoders (\ie the online encoder and the target encoder) that are optimized and updated iteratively.
The online encoder is optimized towards the target encoder, while the target encoder is a momentum encoder. Its parameters are updated as an exponentially moving average of the online encoder.
BYOL uses both a predictor on the online encoder and a ``stop gradient'' operator on the target encoder to avoid network collapse.
SimSiam verifies that a ``stop gradient'' operator is crucial for preventing collapse.
However, it shares the parameters between the online and target encoders.
On the contrary, Barlow Twins~\cite{zbontar2021barlow} uses a 
symmetric architecture by designing an innovative objective function that can align the cross-correlation matrix computed from two contrastive representations as close to the identity matrix as possible.  

Derived from BYOL, the recently proposed self-supervised framework, BUIR~\cite{lee2021bootstrapping}, learns the representations of users and items solely from positive interactions. 
It introduces different views and leverages a slow-moving average network to update the parameters of the target encoder with the online encoder.
Same inputs are fed into different but relevant encoders to generate the contrastive views.

With the booming of SSL in CV and NLP, whether and how multi-modal features can enhance the representations of users and items under the SSL paradigm in recommendation is still unexplored.
In this paper, we propose a simplified yet highly efficient SSL model for multi-modal recommendation. It can achieve outstanding accuracy while also alleviating the computational complexity and memory cost on large graphs.

\section{Bootstrapped Multi-modal Model}
\label{sec:model}
\begin{figure*}[bpt]
	\centering
	\includegraphics[trim=10 10 10 10, clip, width=0.98\textwidth]{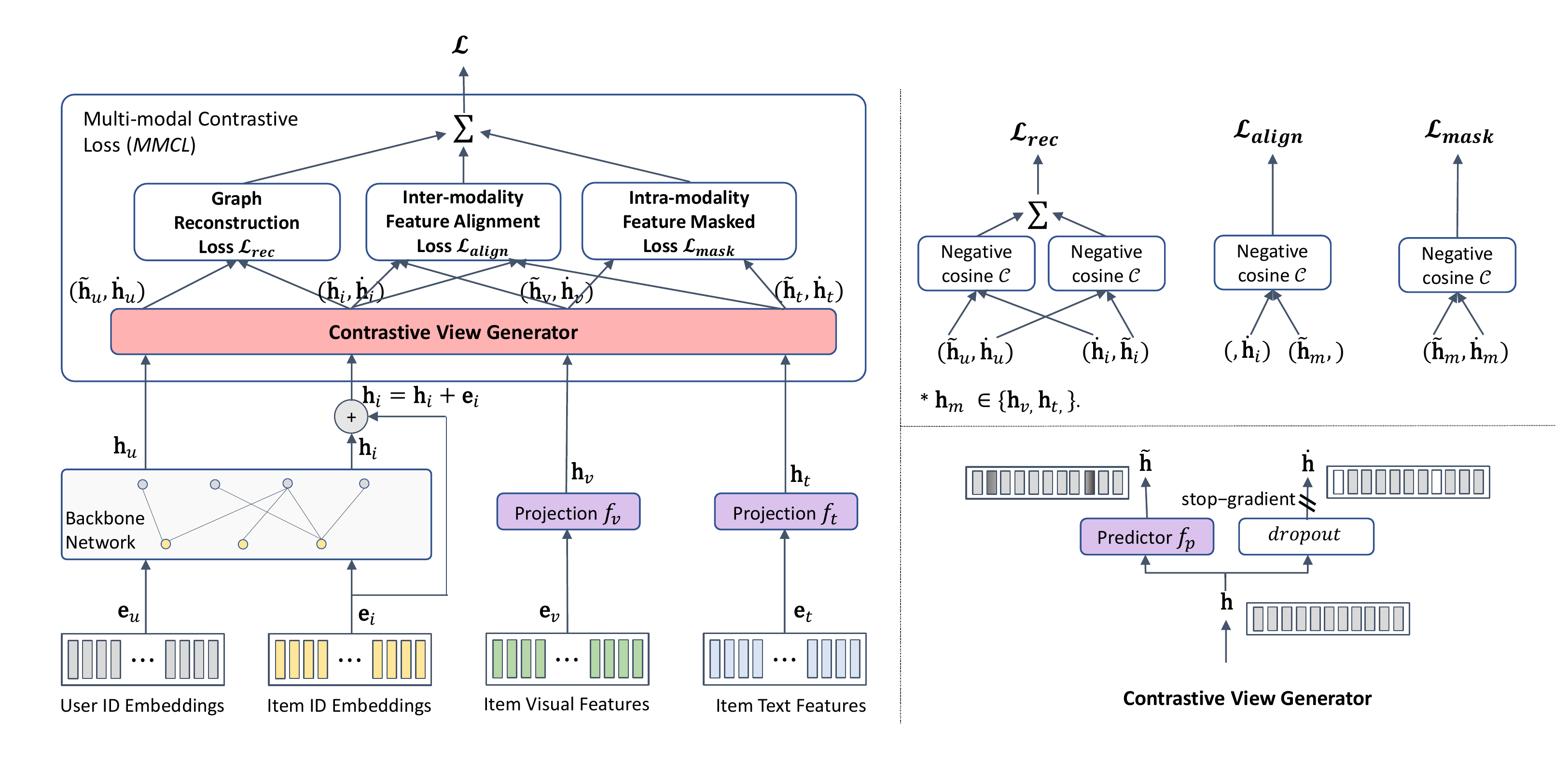}
	\vspace{-10pt}
	\caption{The structure overview of the proposed \bm{}. Projections $f_v$ and $f_t$, as well as predictor $f_p$, are all one-layer MLPs. The parameters of predictor $f_p$ are shared in the Contrastive View Generator (\textit{bottom left}) for ID embeddings and multi-modal latent representations.}
	\Description{The structure overview of the proposed \bm{}. }
	\vspace{-10pt}
	\label{fig:bm3}
\end{figure*}

In this section, we elaborate on our bootstrapped multi-modal model, which encompasses three components as illustrated in Fig.~\ref{fig:bm3}: a) multi-modal latent space convertor, b) contrastive view generator, and c) multi-modal contrastive loss.

\subsection{Multi-modal Latent Space Convertor}
Let $\mathbf{e}_u, \mathbf{e}_i \in \mathbb{R}^d$ denote the input ID embeddings of the user $u \in \mathcal{U} $ and the item $i \in \mathcal{I} $, where $d$ is the embedding dimension, and $\mathcal{U}, \mathcal{I}$ are the sets of users and items, respectively.
Their cardinal numbers are set as $|\mathcal{U}|$ and $|\mathcal{I}|$, respectively.
We denote the modality-specific features obtained from the pre-trained model as $\mathbf{e}_m \in \mathbb{R}^{d_m}$, where $m\in \mathcal{M}$ denotes a specific modality from the full set of modalities $\mathcal{M}$, and $d_m$ denotes the dimension of the features. The cardinal number of $\mathcal{M}$ is denoted by $|\mathcal{M}|$.
In this paper, we consider two modalities: vision $v$ and text $t$.
However, the model can be easily extended to scenarios with more than two modalities.
As the multi-modal feature spaces are different from each other, we first convert the multi-modal features and ID embeddings into the same latent space.

\subsubsection{Multi-modal Features} 
The features of an item obtained from different modalities are of different dimensions and in different feature spaces.
For a multi-modal feature vector $\mathbf{e}_m$, we first project it into a latent low dimension using a projection function $f_m$ based on Multi-Layer Perceptron (MLP).
Then, we have:
\begin{equation}
	\mathbf{h}_m = \mathbf{e}_m \mathbf{W}_m  + \mathbf{b}_m,
\end{equation}
where $\mathbf{W}_m \in \mathbb{R}^{d_m \times d}, \mathbf{b}_m \in \mathbb{R}^d$ denote the linear transformation matrix and bias in the MLP of $f_m$.
In this way, each uni-modal latent representation $\mathbf{h}_m$ shares the same space with ID embeddings.

\subsubsection{ID Embeddings} 
Previous work~\cite{zhang2021mining} has verified the crucial role of ID embeddings in multi-modal recommendation. 
Although the ID embeddings of users and items can be directly initialized within the latent space, they do not encode any structural information about the user-item interaction graph.
Inspired by the recent success of applying GCN for recommendation, we use a backbone network of LightGCN~\cite{he2020lightgcn} with residual connection to encode the structure of the user-item interaction graph.

Suppose $\mathcal{G} = (\mathcal{V},\mathcal{E})$ be a given graph with node set $\mathcal{V}=\mathcal{U} \cup \mathcal{I}$ and edge set $\mathcal{E}$. The number of nodes is denoted by $|\mathcal{V}|$, and the number of edges is denoted by $|\mathcal{E}|$. 
The adjacency matrix is denoted by $\mathbf{A} \in \mathbb{R}^{|\mathcal{V}|\times |\mathcal{V}|}$, and the diagonal degree matrix is denoted by $\mathbf{D}$.
In $\mathcal{G}$, the edges describe observed user-item interactions. If a user has interactions with an item, we build an edge between the user node and the item node.
Moreover, we use $\mathbf{H}^l \in \mathbb{R}^{|\mathcal{V}| \times d}$ to denote the ID embeddings at the $l$-th layer by stacking all the embeddings of users and items at layer $l$.
Specifically, the initial ID embeddings $\mathbf{H}^0$ is a collection of embeddings $\mathbf{e}_u$ and $\mathbf{e}_i$ from all users and items.
A typical feed forward propagation GCN~~\cite{kipf2017semi} to calculate the hidden ID embedding $\mathbf{H}^{l+1}$ at layer $l+1$ is recursively conducted as:
\begin{equation}
	\mathbf{H}^{l+1} = \sigma\left( \hat{\mathbf{A}}\mathbf{H}^{l}\mathbf{W}^{l} \right),
	\label{eq:gcn_va}
\end{equation}
where $\sigma(\cdot)$ is a non-linear function, \eg the ReLu function, $\hat{\mathbf{A}} = \hat{\mathbf{D}}^{-1/2}(\mathbf{A}+\mathbf{I})\hat{\mathbf{D}}^{-1/2}$ is the re-normalization of the adjacency matrix $\mathbf{A}$, and $\hat{\mathbf{D}}$ is the diagonal degree matrix of $\mathbf{A}+\mathbf{I}$.
For node classification, the last layer of a GCN is used to predict the label of a node via a $softmax$ classifier.

On top of the vanilla GCN, LightGCN simplifies its structure by removing the feature transformation $\mathbf{W}^l$ and non-linear activation $\sigma(\cdot)$ layers for recommendation.
As they found these two layers impose adverse effects on recommendation performance.
The simplified graph convolutional layer in LightGCN is defined as:
\begin{equation}
	\mathbf{H}^{l+1} = (\mathbf{D}^{-1/2} \mathbf{A} \mathbf{D}^{-1/2})\mathbf{H}^{l},
	\label{eq:lightGCN_p}
\end{equation}
where the node embeddings of the $(l+1)$-th hidden layer are only linearly aggregated from the $l$-th layer with a transition matrix  $\mathbf{D}^{-1/2} \mathbf{A} \mathbf{D}^{-1/2}$.
The transition matrix is exactly the weighted adjacency matrix mentioned above.

We use a readout function to aggregate all representations in hidden layers for user and item final representations.
However, GCNs may suffer from the over-smoothing problem~\cite{liu2020towards, chen2020simple, li2021training}.
Following LATTICE~\cite{zhang2021mining}, we also add a residual connection~\cite{kipf2017semi, chen2020simple} to the item initial embeddings $\mathbf{H}^0_i$ to obtain the final representations of items.
That is:
\begin{equation}
	\begin{split}
		\mathbf{H_u} &= \text{R{\scriptsize EADOUT}}(\mathbf{H}_u^0, \mathbf{H}_u^1, \mathbf{H}_u^{2}, ... ,\mathbf{H}_u^{L}); \\
		\mathbf{H_i} &= \text{R{\scriptsize EADOUT}}(\mathbf{H}_i^0, \mathbf{H}_i^1, \mathbf{H}_i^{2}, ... ,\mathbf{H}_i^{L}) + \mathbf{H}^0_i,
		\label{eq:lgn_layer_update}
	\end{split}
\end{equation}
where the R{\scriptsize EADOUT} function can be any differentiable function. We use the default mean function of LightGCN for its final ID embedding updating.

With the multi-modal latent space convertor, we can obtain three types of latent embeddings: user ID embeddings, item ID embeddings, and uni-modal item embeddings.
In the following section, we illustrate the design of losses in \bm{} for efficient parameter optimization without negative samples.

\subsection{Multi-modal Contrastive Loss}
Previous studies on SSL use the stop-gradient strategy to prevent the model from resulting in a trivial constant solution~\cite{grill2020bootstrap, chen2021exploring}.
Besides, they use online and target networks to make the model parameters learn in a teacher-student manner~\cite{tarvainen2017mean}.
\bm{} simplifies the current SSL paradigm~\cite{grill2020bootstrap, chen2021exploring} by postponing the data augmentation after the encoding of the online network.
We first illustrate the data augmentation in \bm{}.

\begin{table*}[btp]
	\centering
	\def\arraystretch{0.9}
	\setlength\tabcolsep{2.0pt}  
	\caption{Comparison of computational complexity on graph-based multi-modal methods.}
	\begin{center}
		\begin{tabular}{l|c|c|c|c|c}
			\hline
			\textbf{Component} & \textbf{MMGCN} & \textbf{GRCN} & \textbf{DualGNN} & \textbf{LATTICE} & \textbf{\bm{}} \\
			\hline
			Graph Convolution & $\mathcal{O}(|\mathcal{M}| X)$  & $\mathcal{O}((|\mathcal{M}|+1) X)$ & $\mathcal{O}(|\mathcal{M}| X + kd|\mathcal{U}|)$ & $\mathcal{O}(X)$ & $\mathcal{O}(X)$ \\
			Feature Transform  & $\mathcal{O}(\displaystyle \sum_{m \in \mathcal{M}} |\mathcal{I}|(d_m + d)d_h)$ & $\mathcal{O}(\displaystyle \sum_{m \in \mathcal{M}} |\mathcal{I}|d_m d)$ & $\mathcal{O}(\displaystyle \sum_{m \in \mathcal{M}} |\mathcal{I}|(d_m + d)d_h)$ & \begin{tabular}{@{}c@{}}$\mathcal{O}(|\mathcal{I}|^3 + \displaystyle \sum_{m \in \mathcal{M}} |\mathcal{I}|^2 d_m$ \\ $+ k|\mathcal{I}|\log(|\mathcal{I}|))$ \end{tabular}  & $\mathcal{O}(\displaystyle \sum_{m \in \mathcal{M}} |\mathcal{I}|d_m d)$\\
			BPR/CL Losses & $\mathcal{O}(2dB)$ & $\mathcal{O}((2+|\mathcal{M}|)d B)$ & $\mathcal{O}((2+|\mathcal{M}|)d B)$ & $\mathcal{O}(2d B)$ & $\mathcal{O}((2 + 2|\mathcal{M}|)d B)$ \\
			\hline
			\multicolumn{6}{l}{To fit the page, we set $X = 2L |\mathcal{E}| d/ B$, and $d_h$ denotes the dimension of the hidden layer in a two-layer MLP.}
		\end{tabular}
		\vspace{-10pt}
		\label{tab:com_com}	
	\end{center}
\end{table*}

\subsubsection{Contrastive View Generator}
Prior studies~\cite{lee2021bootstrapping, wu2020self, thakoor2021large} use graph augmentations to generate two alternate views of the original graph for self-supervised learning.
Input features are encoded through both graphs to generate the contrastive views.
To reduce the computational complexity and the memory cost, \bm{} removes the requirement of graph augmentations with a simple latent embedding dropout technique that is analogous to node dropout~\cite{srivastava2014dropout}.
The contrastive latent embedding $\dot{\mathbf{h}}$ of $\mathbf{h}$ under a dropout ratio $p$ is calculated as:
\begin{equation}
	\dot{\mathbf{h}} = \mathbf{h} \cdot \text{Bernoulli}(p).
\end{equation}

Following~\cite{grill2020bootstrap, chen2021exploring}, we also place stop-gradient on the contrastive view $\dot{\mathbf{h}}$. Whilst we feed the original embedding  $\mathbf{h}$ into a predictor of MLP.
\begin{equation}
	\tilde{\mathbf{h}} = \mathbf{h} \mathbf{W}_p  + \mathbf{b}_p,
\end{equation}
where $\mathbf{W}_p \in \mathbb{R}^{d \times d}, \mathbf{b}_p \in \mathbb{R}^d$ denote the linear transformation matrix and bias in the predictor function $f_p$.

\subsubsection{Graph Reconstruction Loss}
\bm{} takes a positive user-item pair $(u,i)$ as input. 
With the generated contrastive view $(\dot{\mathbf{h}}_u, \dot{\mathbf{h}}_i)$ of the online representations $(\tilde{\mathbf{h}}_u, \tilde{\mathbf{h}}_i)$, we define a symmetrized loss function as the negative cosine similarity between $(\tilde{\mathbf{h}}_u, \dot{\mathbf{h}}_i)$ and $(\dot{\mathbf{h}}_u, \tilde{\mathbf{h}}_i)$:
\begin{equation}
	\mathcal{L}_{rec} = \mathcal{C}(\tilde{\mathbf{h}}_u, \dot{\mathbf{h}}_i) + \mathcal{C}(\dot{\mathbf{h}}_u, \tilde{\mathbf{h}}_i).
	\label{eq:loss}
\end{equation}
Function $\mathcal{C}(\cdot, \cdot)$ in the above equation is defined as:
\begin{equation}
	\mathcal{C}(\mathbf{h}_u, \mathbf{h}_i) = -\frac{\mathbf{h}_u^T \mathbf{h}_i}{||\mathbf{h}_u||_2 ||\mathbf{h}_i||_2},
\end{equation}
where $||\cdot||_2$ is $\ell_2$-norm.
The total loss is averaged over all user-item pairs.
The intuition behind this is that we intend to maximize the prediction of the positively perturbed item $i$ given a user $u$, and vice versa. The minimized possible value for this loss is $-1$.

Finally, we stop gradient on the target network and force the backpropagation of loss over the online network only.
We follow the stop gradient ($sg$) operator as in~\cite{grill2020bootstrap, chen2021exploring}, and implement the operator by updating Eq.~\eqref{eq:loss} as:
\begin{equation}
	\mathcal{L}_{rec} = \mathcal{C}(\tilde{\mathbf{h}}_u, sg(\dot{\mathbf{h}}_i)) + \mathcal{C}(sg(\dot{\mathbf{h}}_u), \tilde{\mathbf{h}}_i).
	\label{eq:loss_final}
\end{equation}
With the stop gradient operator, the target network receives no gradient from $(\dot{\mathbf{h}}_u, \dot{\mathbf{h}}_i)$.

\subsubsection{Inter-modality Feature Alignment Loss}
In addition, we further align the multi-modal features of items with their target ID embeddings.
The alignment encourages the ID embeddings close to each other on items with similar multi-modal features. 
For each uni-modal latent embedding $\mathbf{h}_m$ of an item $i$, the contrastive view generator outputs its contrastive pair as $(\tilde{\mathbf{h}}_m^i, \dot{\mathbf{h}}_m^i)$.
We use the negative cosine similarity to perform the alignment between $\dot{\mathbf{h}}_i$ and $\tilde{\mathbf{h}}_m^i$: 
\begin{equation}
	\mathcal{L}_{align} = \mathcal{C}(\tilde{\mathbf{h}}_m^i, \dot{\mathbf{h}}_i).
	\label{eq:inter_loss}
\end{equation}

\subsubsection{Intra-modality Feature Masked Loss}
Finally, \bm{} uses intra-modlity feature masked loss to further encourage the learning of predictor with sparse representations of latent embeddings. Sparse is verified scale efficient in large transformers~\cite{jaszczur2021sparse, shi2022visual}.
We randomly mask out a subset of the latent embedding $\mathbf{h}_m$ by dropout with the contrastive view generator and denote the sparse embedding as $\dot{\mathbf{h}}_m^i$.
The intra-modality feature masked loss is defined as:
\begin{equation}
	\mathcal{L}_{mask} = \mathcal{C}(\tilde{\mathbf{h}}_m^i, \dot{\mathbf{h}}_m^i).
	\label{eq:intra_loss}
\end{equation}

Additionally, we add regularization penalty on the online embeddings (i.e., $\mathbf{h}_u$ and $\mathbf{h}_i$).
Our final loss function is:
\begin{equation}
	\mathcal{L} = \mathcal{L}_{rec} + \mathcal{L}_{align} +  \mathcal{L}_{mask} + \lambda \cdot (||\mathbf{h}_u||^2_2 + ||\mathbf{h}_i||^2_2).
	\label{eq:final_loss}
\end{equation}

\subsection{Top-$K$ Recommendation}
To generate item recommendations for a user, we first predict the interaction scores between the user and candidate items. Then, we rank candidate items based on the predicted interaction scores in descending order, and choose $K$ top-ranked items as recommendations to the user. 
Classical CF methods recommend top-$K$ items by ranking scores of the inner product of a user embedding with all candidate item embeddings.
As our MMCL can learn a good predictor on user and item latent embeddings, we use the embeddings transformed by the predictor $f_p$ for inner product.
That is:
\begin{equation}
	s(\mathbf{h}_u, \mathbf{h}_i) = \tilde{\mathbf{h}}_u \cdot \tilde{\mathbf{h}}_i^T.
\end{equation}
A high score suggests that the user prefers the item.

\subsection{Computational Complexity}
The computational cost of \bm{} mainly occurs in linear propagation of the normalized adjacency matrix $\hat{\mathbf{A}}$.
The analytical complexity of LightGCN and \bm{} are in the same magnitude with $\mathcal{O}(2L |\mathcal{E}| d/B)$ on graph convolutional operation, where $L$ is the number of LightGCN layers, and $B$ is the training batch size.
However, \bm{} has additional costs on multi-modal feature projection and prediction.
The projection cost is $\mathcal{O}(\sum_{m \in \mathcal{M}} |\mathcal{I}|d_m d)$ on all modalities. 
The contrastive loss cost is $\mathcal{O} ((2+2|\mathcal{M}|) d B)$.
The total computational cost for \bm{} is $\mathcal{O}(2L |\mathcal{E}| d/ B + \sum_{m \in \mathcal{M}} |\mathcal{I}|d_m d + (2+2|\mathcal{M}|) d B)$. 
We summarize the computational complexity of the graph-based multi-modal methods in Table~\ref{tab:com_com}.
Both MMGCN and DualGNN use a two-layer MLP for multi-modal feature projection.
On the contrary, LATTICE constructs an item-item graph from the multi-modal features. It costs $\mathcal{O}(|\mathcal{I}|^2 d_m)$ to build the similarity matrix between items, $\mathcal{O}(|\mathcal{I}|^3)$ to normalize the matrix, and $\mathcal{O}(k |\mathcal{I}| \log(|\mathcal{I}|))$ to retrieve top-$k$ most similar items for each item.

\begin{table}[tbp]
	\centering	
	\def\arraystretch{0.9}
	\caption{Statistics of the experimental datasets.}
	\begin{center}
		\begin{tabular}{l|c|c|c|c}
			\hline
			\textbf{Datasets} & \textbf{\# Users} & \textbf{\# Items} & \textbf{\# Interactions} & \textbf{Sparsity} \\
			\hline
			Baby & 19,445 & 7,050 & 160,792 & 99.88\% \\
			Sports & 35,598 & 18,357 & 296,337 & 99.95\%\\
			Electronics & 192,403 & 63,001 & 1,689,188 &  99.99\%\\
			\hline
		\end{tabular}
		\label{tab:dataset}	
	\end{center}
\end{table}

\section{Experiments}
\label{sec: experiments}
We perform comprehensive experiments to evaluate the effectiveness and efficiency of \bm{} to answer the following research questions.
\begin{itemize}
	\item{\textbf{RQ1}}: Can the self-supervised model leveraging only positive user-item interactions outperform or match the performance of the supervised baselines?
	
	\item{\textbf{RQ2}}: How efficient of the proposed \bm{} model in multi-modal recommendation with regard to the computational complexity and memory cost?
	
	\item{\textbf{RQ3}}: To what extent the multi-modal features could affect the recommendation performance of \bm{}?
	
	\item{\textbf{RQ4}}: How different losses in \bm{} affect its recommendation accuracy?
\end{itemize}

\subsection{Experimental Datasets}
Following previous studies~\cite{he2016vbpr,zhang2021mining}, we use the Amazon review dataset~\cite{he2016ups} for experimental evaluation. This dataset provides both product descriptions and images simultaneously, and it is publicly available and varies in size under different product categories. To ensure as many baselines can be evaluated on large-scale datasets, we choose three per-category datasets, \ie Baby, Sports and Outdoors (denoted by \textit{Sports}), and Electronics, for performance evaluation~\footnote{Datasets are available at http://jmcauley.ucsd.edu/data/amazon/links.html}. 
In these datasets, each review rating is treated as a record of positive user-item interaction. This setting has been widely used in previous studies~\cite{he2016vbpr,zhang2021mining,he2020lightgcn,zhang2022diffusion}.
The raw data of each dataset are pre-processed with a 5-core setting on both items and users, and their 5-core filtered results are presented in Table~\ref{tab:dataset}, where the data sparsity is measured as the number of interactions divided by the product of the number of users and the number of items. 
The pre-processed datasets include both visual and textual modalities. Following~\cite{zhang2021mining}, we use the 4,096-dimensional visual features that have been extracted and published in~\cite{ni2019justifying}. For the textual modality, we extract textual embeddings by concatenating the title, descriptions, categories, and brand of each item and utilize sentence-transformers~\cite{reimers2019sentence} to obtain 384-dimensional sentence embeddings.

\begin{table*}[bpt]
	\centering	
	\def\arraystretch{0.9}
	\caption{Overall performance achieved by different recommendation methods in terms of Recall and NDCG.  We mark the global best results on each dataset under each metric in \textbf{boldface} and the second best is \underline{underlined}. }
	\begin{center}
		\begin{tabular}{l|l|ccc|cccccc}
			\hline
			\multirow{2}{*}{\textbf{Datasets}} & \multirow{2}{*}{\textbf{Metrics}} & \multicolumn{3}{c}{\textbf{General models}} & \multicolumn{6}{|c}{\textbf{Multi-modal models}} \\
			\cline{3-11}
			& & \textbf{BPR} & \textbf{LightGCN} & \textbf{BUIR} & \textbf{VBPR} & \textbf{MMGCN} & \textbf{GRCN} & \textbf{DualGNN} & \textbf{LATTICE} & \textbf{\bm} \\
			\hline
			\multirow{4}{*}{Baby} & R@10 & 0.0357 & 0.0479 & 0.0506 & 0.0423 & 0.0378 & 0.0532 & 0.0448 & \underline{0.0544} & \textbf{0.0564}\\
			& R@20 & 0.0575 & 0.0754 & 0.0788 & 0.0663 & 0.0615 & 0.0824 & 0.0716 & \underline{0.0848} & \textbf{0.0883} \\
			& N@10 & 0.0192 & 0.0257 & 0.0269 & 0.0223 & 0.0200 & 0.0282 & 0.0240 & \underline{0.0291} & \textbf{0.0301}  \\
			& N@20 & 0.0249 & 0.0328 & 0.0342 & 0.0284 & 0.0261 & 0.0358 & 0.0309 & \underline{0.0369} & \textbf{0.0383}  \\
			\hline
			\multirow{4}{*}{Sports} & R@10 & 0.0432 & 0.0569 & 0.0467 & 0.0558 & 0.0370 & 0.0559 & 0.0568 & \underline{0.0618} & \textbf{0.0656}\\
			& R@20 & 0.0653 & 0.0864 & 0.0733 & 0.0856 & 0.0605 & 0.0877 & 0.0859 & \underline{0.0947} & \textbf{0.0980} \\
			& N@10 & 0.0241 & 0.0311 & 0.0260 & 0.0307 & 0.0193 & 0.0306 & 0.0310 & \underline{0.0337} & \textbf{0.0355}  \\
			& N@20 & 0.0298 & 0.0387 & 0.0329 & 0.0384 & 0.0254 & 0.0389 & 0.0385 & \underline{0.0422} & \textbf{0.0438}  \\
			\hline
			\multirow{4}{*}{Electronics} & R@10 & 0.0235 & \underline{0.0363} & 0.0332 & 0.0293 & 0.0207 & 0.0349 & \underline{0.0363} & - & \textbf{0.0437}\\
			& R@20 & 0.0367 & 0.0540 & 0.0514 & 0.0458 & 0.0331 & 0.0529 & \underline{0.0541} & - & \textbf{0.0648} \\
			& N@10 & 0.0127 & \underline{0.0204} & 0.0185 & 0.0159 & 0.0109 & 0.0195 & 0.0202 & - & \textbf{0.0247}  \\
			& N@20 & 0.0161 & \underline{0.0250} & 0.0232 & 0.0202 & 0.0141 & 0.0241 & 0.0248 & - & \textbf{0.0302}  \\
			\hline			
			\multicolumn{11}{l}{`-' indicates the model cannot be fitted into a Tesla V100 GPU card with 32 GB memory.}
		\end{tabular}
		\vspace{-10pt}
		\label{tab:perform}	
	\end{center}
\end{table*}

\subsection{Baseline Methods}
To demonstrate the effectiveness of \bm, we compare it with the following state-of-the-art recommendation methods, including general CF recommendation models and multi-modal recommendation models.
\begin{itemize}
	\item \textbf{BPR}~\cite{rendle2009bpr}: This is a matrix factorization model optimized by a pair-wise ranking loss in a Bayesian way.
	\item \textbf{LightGCN}~\cite{he2020lightgcn}: This is a simplified graph convolution network that only performs linear propagation and aggregation between neighbors. The hidden layer embeddings are averaged to calculate the final user and item embeddings for prediction.
	\item \textbf{BUIR}~\cite{lee2021bootstrapping}: This self-supervised framework uses asymmetric network architecture to update its backbone network parameters. In BUIR, LightGCN is used as the backbone network. 
	It is worth noting that BUIR does not rely on negative samples for learning.
	\item \textbf{VBPR}~\cite{he2016vbpr}: This model incorporates visual features for user preference learning with BPR loss. Following~\cite{zhang2021mining, wang2021dualgnn}, we concatenate the multi-modal features of an item as its visual feature for user preference learning.
	\item \textbf{MMGCN}~\cite{wei2019mmgcn}: This method constructs a modal-specific graph to learn user preference on each modality leveraging GCN. The final user and item representations are generated by combining the learned representations from each modality.
	\item \textbf{GRCN}~\cite{wei2020graph}: This method improves previous GCN-based models by refining the user-item bipartite graph with removal of false-positive edges. User and item representations are learned on the refined bipartite graph by performing information propagation and aggregation.
	\item \textbf{DualGNN}~\cite{wang2021dualgnn}: This method builds an additional user-user correlation graph from the user-item bipartite graph and uses it to fuse the user representation from its neighbors in the correlation graph.
	\item \textbf{LATTICE}~\cite{zhang2021mining}: This method mines the latent structure between items by learning an item-item graph from their multi-modal features. Graph convolutional operations are performed on both item-item graph and user-item interaction graph to learn user and item representations.
\end{itemize}

We group the first three baselines (\ie BPR, LightGCN, and BUIR) as general models, because they only use implicit feedback (\ie user-item interactions) for recommendation. The other multi-modal models utilize both implicit feedback and multi-modal features for recommendation. Analogously, we categorize BUIR into the self-supervised model and the others as supervised models as they are using negative samples for representation learning.
The proposed \bm{} model is within the self-supervised multi-modal domain.

\subsection{Setup and Evaluation Metrics}
For a fair comparison, we follow the same evaluation setting of~\cite{zhang2021mining, wang2021dualgnn} with a random data splitting 8:1:1 on the interaction history of each user for training, validation and testing. 
Moreover, we use $\text{Recall}@K$ and $\text{NDCG}@K$ to evaluate the top-$K$ recommendation performance of different recommendation methods. Specifically, we use the all-ranking protocol instead of the negative-sampling protocol to compute the evaluation metrics for recommendation accuracy comparison. In the recommendation phase, all items that have not been interacted by the given user are regarded as candidate items. 
In the experiments, we empirically report the results of $K$ at 10, 20 and abbreviate the metrics of $\text{Recall}@K$ and $\text{NDCG}@K$ as $\text{R}@K$ and $\text{N}@K$, respectively.

\subsection{Implementation Details}
Same as other existing work~\cite{he2020lightgcn, zhang2021mining}, we fix the embedding size of both users and items to 64 for all models, initialize the embedding parameters with the Xavier method~\cite{glorot2010understanding}, and use Adam~\cite{kingma2015adam} as the optimizer with a learning rate of 0.001. For a fair comparison, we carefully tune the parameters of each model following their published papers. 
The proposed \bm{} model is implemented by PyTorch~\cite{paszke2019pytorch}. We perform a grid search across all datasets to conform to its optimal settings. Specifically, the number of GCN layers is tuned in \{1, 2\}. The dropout rate for embedding perturbation is chosen from \{0.3, 0.5\}, and the regularization coefficient is searched in \{0.1, 0.01\}.
For convergence consideration, the early stopping and total epochs are fixed at 20 and 1000, respectively. 
Following~\cite{zhang2021mining}, we use R@20 on the validation data as the training stopping indicator.
We have integrated our model and all baselines into the unified multi-modal recommendation platform, MMRec~\cite{zhou2023mmrecsm}.

\begin{table*}[tbp]
	\centering	
	\def\arraystretch{0.9}
	\setlength\tabcolsep{5.0pt}
	\caption{Efficiency comparison of \bm{} against the baselines.}
	\begin{center}
		\begin{tabular}{l|l|ccc|cccccc}
			\hline
			\multirow{2}{*}{\textbf{Datasets}} & \multirow{2}{*}{\textbf{Metrics}} & \multicolumn{3}{c}{\textbf{General models}} & \multicolumn{6}{|c}{\textbf{Multi-modal models}} \\
			\cline{3-11}
			& & \textbf{BPR} & \textbf{LightGCN} & \textbf{BUIR} & \textbf{VBPR} & \textbf{MMGCN} & \textbf{GRCN} & \textbf{DualGNN*} & \textbf{LATTICE} & \textbf{\bm} \\
			\hline
			\multirow{2}{*}{Baby} & Memory (GB) & 1.59 & 1.69 & 2.29 & 1.89 & 2.69 & 2.95 & 2.05 & 4.53 & 2.11\\
			& Time ($s$/epoch) & 0.47 & 0.99 & 0.77 & 0.57 & 3.48 & 2.36 & 7.81 & 1.61 & 0.85 \\
			\hline
			\multirow{2}{*}{Sports} & Memory (GB) & 2.00 & 2.24 & 3.75 & 2.71 & 3.91 & 4.49 & 2.81 & 19.93 & 3.58\\
			& Time ($s$/epoch) & 0.95 & 2.86 & 2.19 & 1.28 & 16.60 & 6.74 & 12.60 & 10.71 & 3.03 \\
			\hline
			\multirow{2}{*}{Electronics} & Memory (GB) & 3.69 & 4.92 & 10.13 & 6.20 & 14.54 & 17.38 & 8.85 & - & 8.28 \\
			& Time ($s$/epoch) & 6.75 & 67.49 & 63.77 & 14.20 & 470.15 & 152.68 & 341.02 & - & 73.31 \\
			\hline
			\multicolumn{11}{l}{`-' denotes the model cannot be fitted into a Tesla V100 GPU card with 32 GB memory. }\\
			\multicolumn{11}{l}{`*' In pre-processing, DualGNN requires about 138GB memory and 6 hours to construct the user-user relationship graph on Electronics data.}
		\end{tabular}
		\vspace{-10pt}
		\label{tab:perform_eff}	
	\end{center}
\end{table*}

\subsection{Effectiveness of \bm{} (RQ1)}
The performance achieved by different recommendation methods on all three datasets are summarized in Table~\ref{tab:perform}. From the table, we have the following observations. 
\textit{First,} the proposed \bm{} model significantly outperforms both general recommendation methods and state-of-the-art multi-modal recommendation methods on each dataset. Specifically, \bm{} improves the best baselines by 3.68\%, 6.15\%, and 20.39\% in terms of Recall@10 on Baby, Sports, and Electronics, respectively.
The results not only verify the effectiveness of \bm{} in recommendation, but also show \bm{} is superior to the baselines for recommendation on the large graph (\ie Electronics).
\textit{Second,} multi-modal recommendation models do not always outperform the general recommendation models without leveraging modal features. Although the recommendation accuracy of VBPR building upon BPR dominates its counterpart (\ie BPR) across all datasets, GRCN and DualGNN using LightGCN as its downstream CF model do not gain much improvement over LightGCN. Differing from the multi-modal feature fusion mechanism of MMGCN, GRCN, and DualGNN, LATTICE uses the multi-modal features in an indirect manner by building an item-item relation graph and performs graph convolutional operation on the graph. We speculate there are two potential reasons leading to the suboptimal performance of MMGCN, GRCN, and DualGNN. i). They fuse the item ID embedding with its modal-specific features. Table~\ref{tab:perform} shows that LightGCN with ID embeddings can obtain good recommendation accuracy. The mixing of ID embeddings and modal features causes the items to lose their identities in recommendation, resulting in accuracy degradation. ii). They fail to differentiate the importance of multi-modal features of items. In MMGCN, GRCN, and DualGNN, they treat features from each modality equally. However, our ablation study in Section~\ref{ssec:ablation} shows that the extracted visual features may contain noise and are less informative than the textual features. On the contrary, LATTICE learns the weights between multi-modal features when building the item-item graph. The proposed \bm{} model alleviates the above issues by placing the contributions of multi-modal features into the loss function.
\textit{Finally,} we compare the recommendation accuracy between self-supervised learning models (\ie BUIR and \bm{}). Although BUIR shows comparable performance with LightGCN, it is inferior to \bm{}. The performance of BUIR depends on the perturbed graph. Better contrastive view results in better recommendation acrruacy. As a result, it obtains fluctuating performance over different datasets. \bm{} reduces the requirement of graph augmentation by latent embedding dropout. It is earlier for \bm{} to obtain a consistent contrastive view with the original view than BUIR. Moreover, \bm{} is more efficient than BUIR, because it uses only one backbone network.

\subsection{Efficiency of \bm{} (RQ2)}
Apart from the comparison of recommendation accuracy, we also report the efficiency of \bm{} against the baselines, in terms of utilized memory and training time per epoch. It is worth noting that all models are firstly evaluated on a GeForce RTX 2080 Ti with 12GB memory, and the model will be advanced to a Tesla V100 GPU with 32 GB memory if it cannot be fitted into the 12 GB memory. The efficiencies of different methods are summarized in Table~\ref{tab:perform_eff}. 
From the table, we can have the following two observations.
\textit{First,} from both the general model and multi-modal model perspectives, graph-based models usually consume more memory than classic CF models (\ie BPR and VBPR). 
	Specifically, classic CF models require a minimum GPU memory cost for representation learning of users and items. Whilst graph-based models usually need to retain an additional user-item interaction graph for information propagation and aggregation. Moreover, graph-based multi-modal recommendation models need more memory, as they use both the user-item graph and multi-modal features in general.
\textit{Second,} among the graph-based multi-modal recommendation models, \bm{} consumes less or comparable memory than other baselines. However, it reduces the training time by 2--9$\times$ per epoch.
	Compared with the best baseline, \bm{} requires only half of the training time and half of the consumed memory of LATTICE. 
	Although \bm{} uses LightGCN as its backbone model, it does not introduce much additional cost on LightGCN other than the multi-modal features. The reason is that \bm{} removes the negative sampling time and uses fewer GCN layers. %

\begin{table}[tbp]
	\centering	
	\def\arraystretch{0.9}
	\setlength\tabcolsep{3.0pt}
	\caption{Ablation study of \bm{} on multi-modal features.}
	\vspace{-10pt}
	\begin{center}
		\begin{tabular}{l|l|cccc}
			\hline
			\textbf{Datasets} & \textbf{Variants} & \textbf{R@10} & \textbf{R@20} & \textbf{N@10} & \textbf{N@20} \\
			\hline
			\multirow{4}{*}{Baby} & \bm{}\textsubscript{w/o v\&t} & 0.0506 & 0.0793 & 0.0273 & 0.0347 \\
			& \bm{}\textsubscript{w/o t} & 0.0518 & 0.0820 & 0.0277 & 0.0354 \\
			& \bm{}\textsubscript{w/o v} & 0.0522 & 0.0828 & 0.0279 & 0.0358 \\
			& \bm{} & 0.0564 & 0.0883 & 0.0301 & 0.0383 \\
			\hline
			\multirow{4}{*}{Sports} & \bm{}\textsubscript{w/o v\&t} & 0.0600 & 0.0927 & 0.0326 & 0.0410 \\
			& \bm{}\textsubscript{w/o t} & 0.0641 & 0.0976 & 0.0349 & 0.0435 \\
			& \bm{}\textsubscript{w/o v} & 0.0647 & 0.0968 & 0.0349  & 0.0432  \\
			& \bm{} & 0.0656  & 0.0980 & 0.0355 & 0.0438 \\
			\hline
			\multirow{4}{*}{Elec.} & \bm{}\textsubscript{w/o v\&t} & 0.0427 & 0.0633 & 0.0240 & 0.0293 \\
			& \bm{}\textsubscript{w/o t} & 0.0423 & 0.0632 & 0.0237 & 0.0291 \\
			& \bm{}\textsubscript{w/o v} & 0.0423 & 0.0633 & 0.0236 & 0.0290 \\
			& \bm{} & 0.0437 & 0.0648 & 0.0247 & 0.0302 \\
			\hline
		\end{tabular}
		\vspace{-15pt}
		\label{tab:abla_feat}	
	\end{center}
\end{table}

\subsection{Ablation Study (RQ3 \& RQ4)}
\label{ssec:ablation}

To fully understand the behaviors of \bm{}, we perform ablation studies on both the multi-modal features and different parts of the multi-modal contrastive loss.

\subsubsection{Multi-modal Features (RQ3)}
We evaluate the recommendation accuracy of \bm{} by feeding individual modal features into the model.
Specifically, we design the following variants of \bm{}.
\begin{itemize}
	\item \bm{}\textsubscript{w/o v\&t}: In this variant, \bm{} degrades to a general recommendation model that exploits only the user-item interactions for recommendation.
	\item \bm{}\textsubscript{w/o v}: This variant of \bm{} learns the representations of users and items without the visual features of items.
	\item \bm{}\textsubscript{w/o t}: This variant of \bm{} is trained without the input from the textual features.
\end{itemize}
Table~\ref{tab:abla_feat} summarizes the recommendation performance of \bm{} and its variants, \ie \bm{}\textsubscript{w/o v\&t}, \bm{}\textsubscript{w/o v}, and \bm{}\textsubscript{w/o t}, on all three experimental datasets. 

As shown in Table~\ref{tab:abla_feat}, the importance of textual and visual features varies with datasets. \bm{}\textsubscript{w/o v} leveraging only on the textual features gains slight better recommendation accuracy than \bm{}\textsubscript{w/o t} on Baby dataset. However, on the other datasets, the differences between these two variations are negligible. Moreover, we can observe that the context features from either textual or visual modality can boost the performance of \bm{}\textsubscript{w/o v\&t} on Baby and Sports datasets. However, this statement does not hold under the large dataset, \ie Electronics. By combining both the textual and visual features, \bm{} achieves the best recommendation accuracy on all three datasets.  

Comparing the experimental results in Table~\ref{tab:abla_feat} with that in Table~\ref{tab:perform}, we note that \bm{}\textsubscript{w/o v\&t} without any multi-modal features is competitive with most multi-modal baseline models and superior to the general baseline models. This demonstrates the effectiveness of the self-supervised learning paradigm for top-$K$ item recommendation. It is worth noting that \bm{} uses LightGCN as its backbone network. The best performance of \bm{} is achieved by using 1, 1, and 2 GCN layers on Baby, Sports, and Electronics datasets, respectively. Whilst LightGCN itself requires 4 GCN layers to achieve its best performance on all datasets.

\begin{table}[tbp]
	\centering	
	\def\arraystretch{0.9}
	\setlength\tabcolsep{3.0pt}
	\caption{Ablation study on different parts of the multi-modal contrastive loss of \bm{}.}
	\vspace{-10pt}
	\begin{center}
		\begin{tabular}{l|l|cccc}
			\hline
			\textbf{Datasets} & \textbf{Variants} & \textbf{R@10} & \textbf{R@20} & \textbf{N@10} & \textbf{N@20} \\
			\hline
			\multirow{4}{*}{Baby} & \bm{}\textsubscript{w/o mm} & 0.0506 & 0.0793 & 0.0273 & 0.0347 \\
			& \bm{}\textsubscript{w/o inter} & 0.0542 & 0.0842 & 0.0289 & 0.0366 \\
			& \bm{}\textsubscript{w/o intra} & 0.0526 & 0.0830 & 0.0281 & 0.0360 \\
			& \bm{} & 0.0564 & 0.0883 & 0.0301 & 0.0383 \\
			\hline
			\multirow{4}{*}{Sports} & \bm{}\textsubscript{w/o mm} & 0.0600 & 0.0927 & 0.0326 & 0.0410 \\
			& \bm{}\textsubscript{w/o inter} & 0.0614 & 0.0941 & 0.0336  & 0.0420  \\
			& \bm{}\textsubscript{w/o intra} & 0.0633 & 0.0947 & 0.0344 & 0.0425 \\
			& \bm{} & 0.0656  & 0.0980 & 0.0355 & 0.0438 \\
			\hline
			\multirow{4}{*}{Elec.} & \bm{}\textsubscript{w/o mm} & 0.0427 & 0.0633 & 0.0240 & 0.0293 \\
			& \bm{}\textsubscript{w/o inter} & 0.0393 & 0.0593 & 0.0218 & 0.0270 \\
			& \bm{}\textsubscript{w/o intra} & 0.0410 & 0.0619 & 0.0227 & 0.0281 \\
			& \bm{} & 0.0437 & 0.0648 & 0.0247 & 0.0302 \\
			\hline
		\end{tabular}
		\vspace{-10pt}
		\label{tab:abla_loss}	
	\end{center}
\end{table}

\subsubsection{Multi-modal Contrastive Loss (RQ4)}

As the loss function plays a critical role in learning the model parameters of \bm{} without negative samples. We further study the behaviors of \bm{} by removing different parts of the multi-modal contrastive loss. Specifically, we consider the following variants of \bm{} for experimental evaluation.
\begin{itemize}
	\item \bm{}\textsubscript{w/o mm}: This variant only uses the interaction graph reconstruction loss to train the model parameters, \ie trained without the multi-modal losses. It is worth noting that this variant is equivalent to \bm{}\textsubscript{w/o v\&t}.
	
	\item \bm{}\textsubscript{w/o inter}: In this variant, the representations of users and items are learned without considering the inter-modality alignment loss.
	
	\item \bm{}\textsubscript{w/o intra}: This variant learns the representations of users and items without considering the feature masked loss.
\end{itemize}
Table~\ref{tab:abla_loss} shows the performance achieved by \bm{}, \bm{}\textsubscript{w/o mm}, \bm{}\textsubscript{w/o inter}, and \bm{}\textsubscript{w/o intra} on all three datasets.

From Table~\ref{tab:abla_loss}, we find a similar pattern as that shown in the ablation study on multi-modal features. That is, the multi-modal losses can improve the recommendation accuracy of \bm{} on Baby and Sports datasets. However, \bm{} leveraging either the inter-modality alignment loss or the intra-modality feature masked loss degrades its performance on the Electronics dataset. Moreover, the importance of inter- and intra-modality losses also varies with the datasets.

From the ablation studies on features and the loss function, we find the recommendation accuracy on the large dataset (\ie Electronics) shows a different pattern from that of the small-scale datasets (\ie Baby and Sports). The uni-modal feature or uni-loss function in \bm{} shows no improvement in recommendation accuracy on Electronics dataset. We speculate that the supervised or self-supervised signals on a large dataset already enable \bm{}\textsubscript{w/o mm} to learn good representations of users and items. Adding coarse multi-modal signals to \bm{}\textsubscript{w/o mm} does not help improve the recommendation accuracy.

\section{Conclusion}
\label{sec:conclusion}
This paper proposes a novel self-supervised learning framework, named \bm{}, for multi-modal recommendation. \bm{} removes the requirement of randomly sampled negative examples in modeling the interactions between users and items. 
To generate a contrastive view in self-supervised learning, \bm{} utilizes a simple yet efficient latent embedding dropout mechanism to perturb the original embeddings of users and items. Moreover, a novel learning paradigm based on the multi-modal contrastive loss has also been devised. Specifically,  
the contrastive loss jointly minimizes: a) the reconstruction loss of the user-item interaction graph, b) the alignment loss between ID embeddings of items and their multi-modal features, and c) the masked loss within a modality-specific feature.
We evaluate the proposed \bm{} model on three real-world datasets, including one large-scale dataset, to demonstrate its effectiveness and efficiency in recommendation tasks. The experimental results show that \bm{} achieves significant accuracy improvements over the state-of-the-art multi-modal recommendation methods, while training 2-9$\times$ faster than the baseline methods. 

\begin{acks}
This work was supported by Alibaba Group through Alibaba Innovative Research (AIR) Program and Alibaba-NTU Singapore Joint Research Institute (JRI), Nanyang Technological University, Singapore.
\end{acks}

\section{Appendices}

\subsection{Hyper-parameter Sensitivity Study}

To guide the selection of hyper-parameters of \bm{}, we perform a hyper-parameter sensitivity study with regard to the recommendation accuracy, in terms of Recall@20. 
We use at least two datasets to evaluate the performance of \bm{} under different hyper-parameter settings. Specifically, we consider the following three hyper-parameters, \ie the number of GCN layers $L$, the ratio of embedding dropout, and the regularization coefficient factor $\lambda$.

\begin{figure}[H]
	\centering
	\includegraphics[trim=10 10 10 10, clip, width=0.45\textwidth]{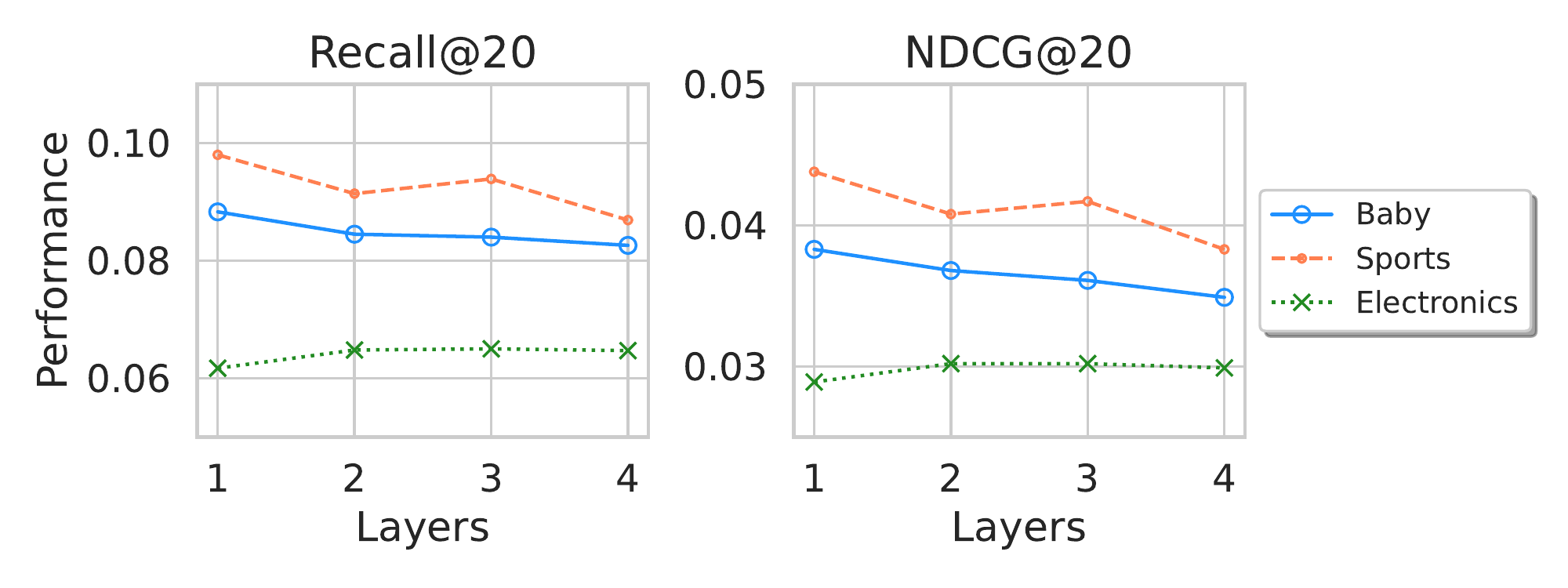}
	\caption{Top-20 recommendation accuracy of \bm{} varies with  the number of GCN layers in the backbone network.}
	\Description{Top-20 recommendation accuracy of \bm{} varies with  the number of GCN layers in the backbone network.}
	\label{fig:layer-20}
\end{figure}

\subsubsection{The Number of GCN Layers}
The number of GCN layers $L$ in \bm{} is varied in \{1, 2, 3, 4\}. Fig.~\ref{fig:layer-20} shows the performance trends of \bm{} with respect to different settings of $L$. As shown in Fig.~\ref{fig:layer-20}, \bm{} shows relatively slow performance degradation as the number of layers increases, on small-scale datasets (\ie Baby and Sports). However, the recommendation accuracy can be improved with more than one GCN layer in the backbone network on Electronics dataset.

\subsubsection{The Dropout Ratio and Regularization Coefficient} 
\begin{figure}[H]
	\centering
	\subfloat[Sports]{\includegraphics[width=0.7\linewidth]{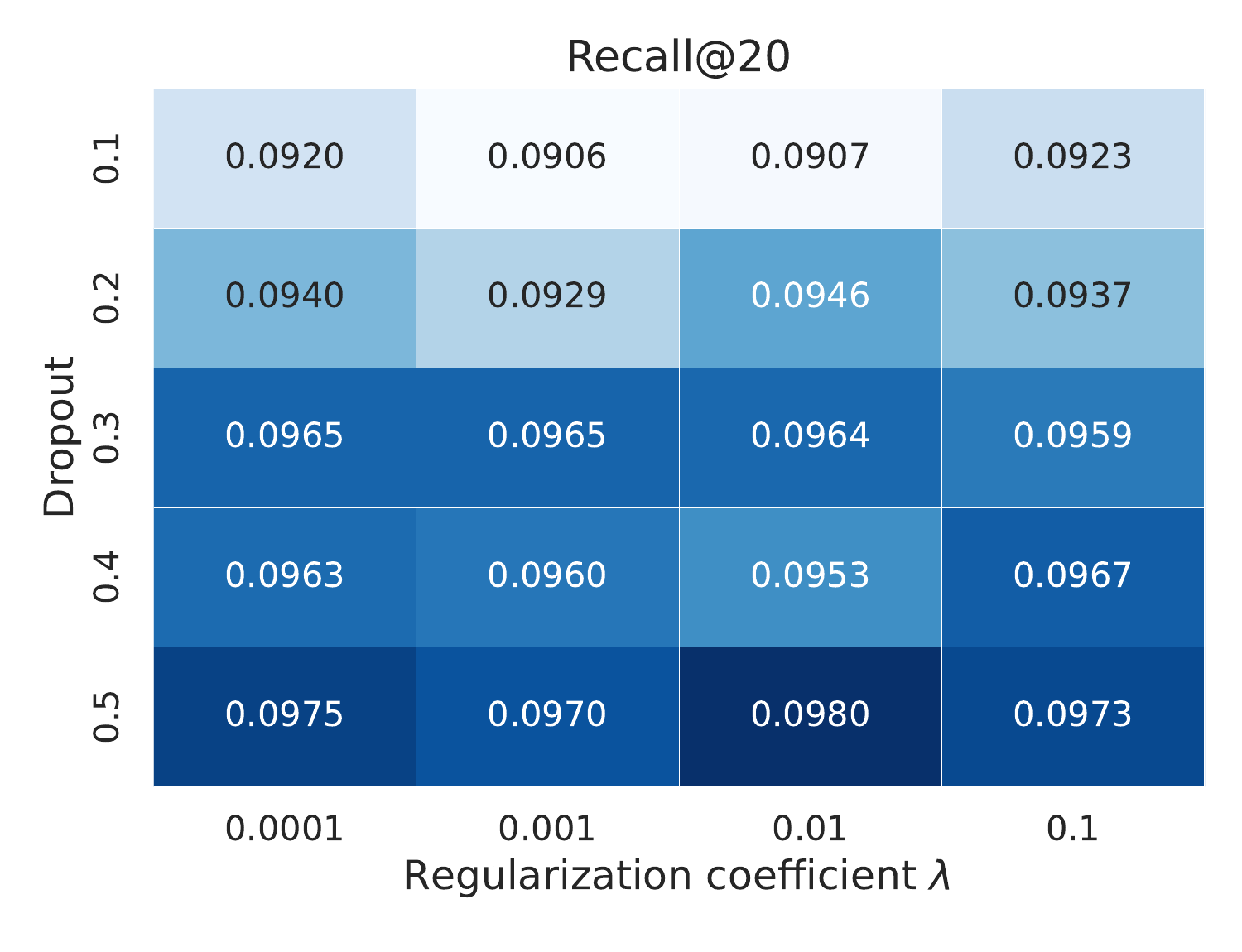}} \hspace{0.1cm}
	\subfloat[Electronics]{\includegraphics[width=0.7\linewidth]{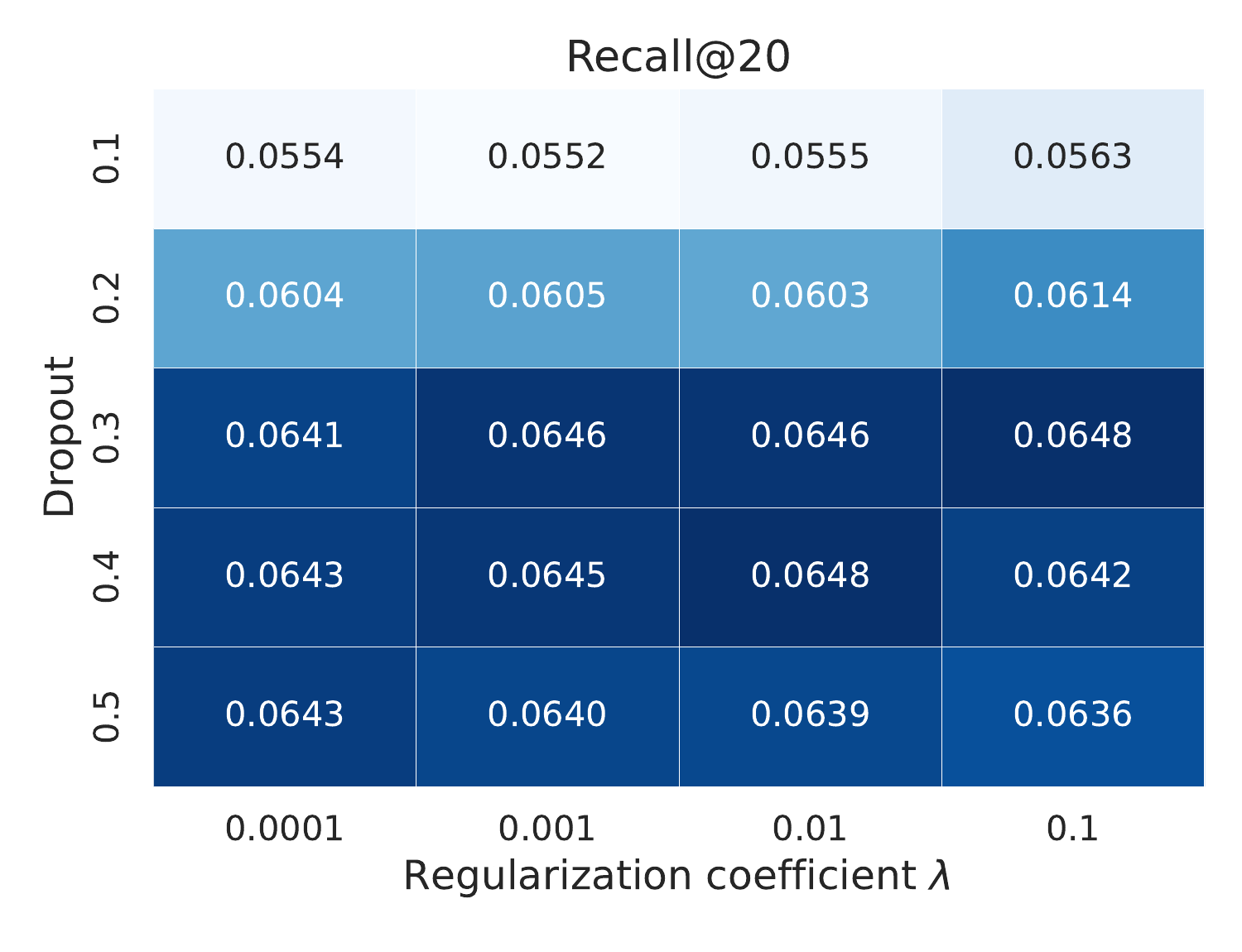}}
	\caption{The Performance achieved by \bm{} with respect to different combinations of the latent embedding dropout ratio and regularization coefficient on three datasets. Darker background indicates better recommendation accuracy.}
	\Description{The Performance achieved by \bm{} with respect to different combinations of the latent embedding dropout ratio and regularization coefficient on three datasets. Darker background indicates better recommendation accuracy.}
	\label{fig:reg-dp-all}
\end{figure}

We vary the dropout ratio of \bm{} from 0.1 to 0.5 with a step of 0.1, and vary the regularization coefficient $\lambda$ in \{0.0001, 0.001, 0.01, 0.1\}. Fig.~\ref{fig:reg-dp-all} shows the performance achieved by \bm{} under different combinations of the embedding dropout ratio and regularization coefficient. We note that a larger dropout ratio of \bm{} on a relative small-scale dataset (\ie Sports) usually helps \bm{} achieve better recommendation accuracy. Moroever, the performance of \bm{} is less sensitive to the settings of regularization coefficient on the large dataset (\ie Electronics). In Fig.~\ref{fig:reg-dp-all}(b), it is worth noting that \bm{} achieves competitive recommendation accuracy when the dropout ratio is larger than 0.2. This verifies the stability of \bm{} in the recommendation task, \ie the recommendation performance of \bm{} is not just a consequence of random seeds.

\bibliographystyle{ACM-Reference-Format}
\bibliography{bm3.bib}


\begin{thebibliography}{48}


\ifx \showCODEN    \undefined \def \showCODEN     #1{\unskip}     \fi
\ifx \showDOI      \undefined \def \showDOI       #1{#1}\fi
\ifx \showISBNx    \undefined \def \showISBNx     #1{\unskip}     \fi
\ifx \showISBNxiii \undefined \def \showISBNxiii  #1{\unskip}     \fi
\ifx \showISSN     \undefined \def \showISSN      #1{\unskip}     \fi
\ifx \showLCCN     \undefined \def \showLCCN      #1{\unskip}     \fi
\ifx \shownote     \undefined \def \shownote      #1{#1}          \fi
\ifx \showarticletitle \undefined \def \showarticletitle #1{#1}   \fi
\ifx \showURL      \undefined \def \showURL       {\relax}        \fi
\providecommand\bibfield[2]{#2}
\providecommand\bibinfo[2]{#2}
\providecommand\natexlab[1]{#1}
\providecommand\showeprint[2][]{arXiv:#2}

\bibitem[Bromley et~al\mbox{.}(1993)]%
        {bromley1993signature}
\bibfield{author}{\bibinfo{person}{Jane Bromley}, \bibinfo{person}{Isabelle
  Guyon}, \bibinfo{person}{Yann LeCun}, \bibinfo{person}{Eduard S{\"a}ckinger},
  {and} \bibinfo{person}{Roopak Shah}.} \bibinfo{year}{1993}\natexlab{}.
\newblock \showarticletitle{Signature verification using a" siamese" time delay
  neural network}.
\newblock \bibinfo{journal}{\emph{Advances in neural information processing
  systems}}  \bibinfo{volume}{6} (\bibinfo{year}{1993}),
  \bibinfo{pages}{737--744}.
\newblock


\bibitem[Chen et~al\mbox{.}(2020b)]%
        {chen2020simple}
\bibfield{author}{\bibinfo{person}{Ming Chen}, \bibinfo{person}{Zhewei Wei},
  \bibinfo{person}{Zengfeng Huang}, \bibinfo{person}{Bolin Ding}, {and}
  \bibinfo{person}{Yaliang Li}.} \bibinfo{year}{2020}\natexlab{b}.
\newblock \showarticletitle{Simple and deep graph convolutional networks}. In
  \bibinfo{booktitle}{\emph{International Conference on Machine Learning}}.
  PMLR, \bibinfo{pages}{1725--1735}.
\newblock


\bibitem[Chen et~al\mbox{.}(2020a)]%
        {chent2020simple}
\bibfield{author}{\bibinfo{person}{Ting Chen}, \bibinfo{person}{Simon
  Kornblith}, \bibinfo{person}{Mohammad Norouzi}, {and}
  \bibinfo{person}{Geoffrey Hinton}.} \bibinfo{year}{2020}\natexlab{a}.
\newblock \showarticletitle{A simple framework for contrastive learning of
  visual representations}. In \bibinfo{booktitle}{\emph{International
  Conference on Machine Learning}}. \bibinfo{pages}{1597--1607}.
\newblock


\bibitem[Chen et~al\mbox{.}(2019)]%
        {chen2019personalized}
\bibfield{author}{\bibinfo{person}{Xu Chen}, \bibinfo{person}{Hanxiong Chen},
  \bibinfo{person}{Hongteng Xu}, \bibinfo{person}{Yongfeng Zhang},
  \bibinfo{person}{Yixin Cao}, \bibinfo{person}{Zheng Qin}, {and}
  \bibinfo{person}{Hongyuan Zha}.} \bibinfo{year}{2019}\natexlab{}.
\newblock \showarticletitle{Personalized fashion recommendation with visual
  explanations based on multimodal attention network: Towards visually
  explainable recommendation}. In \bibinfo{booktitle}{\emph{Proceedings of the
  42nd International ACM SIGIR Conference on Research and Development in
  Information Retrieval}}. \bibinfo{pages}{765--774}.
\newblock


\bibitem[Chen and He(2021)]%
        {chen2021exploring}
\bibfield{author}{\bibinfo{person}{Xinlei Chen} {and} \bibinfo{person}{Kaiming
  He}.} \bibinfo{year}{2021}\natexlab{}.
\newblock \showarticletitle{Exploring Simple Siamese Representation Learning}.
  In \bibinfo{booktitle}{\emph{Proceedings of the IEEE Conference on Computer
  Vision and Pattern Recognition}}. IEEE.
\newblock


\bibitem[Glorot and Bengio(2010)]%
        {glorot2010understanding}
\bibfield{author}{\bibinfo{person}{Xavier Glorot} {and} \bibinfo{person}{Yoshua
  Bengio}.} \bibinfo{year}{2010}\natexlab{}.
\newblock \showarticletitle{Understanding the difficulty of training deep
  feedforward neural networks}. In \bibinfo{booktitle}{\emph{Proceedings of the
  thirteenth international conference on artificial intelligence and
  statistics}}. JMLR Workshop and Conference Proceedings,
  \bibinfo{pages}{249--256}.
\newblock


\bibitem[Grill et~al\mbox{.}(2020)]%
        {grill2020bootstrap}
\bibfield{author}{\bibinfo{person}{Jean-Bastien Grill},
  \bibinfo{person}{Florian Strub}, \bibinfo{person}{Florent Altch{\'e}},
  \bibinfo{person}{Corentin Tallec}, \bibinfo{person}{Pierre~H Richemond},
  \bibinfo{person}{Elena Buchatskaya}, \bibinfo{person}{Carl Doersch},
  \bibinfo{person}{Bernardo~Avila Pires}, \bibinfo{person}{Zhaohan~Daniel Guo},
  \bibinfo{person}{Mohammad~Gheshlaghi Azar}, {et~al\mbox{.}}}
  \bibinfo{year}{2020}\natexlab{}.
\newblock \showarticletitle{Bootstrap your own latent: A new approach to
  self-supervised learning}. In \bibinfo{booktitle}{\emph{Proceedings of the
  34th Annual Conference on Neural Information Processing Systems}}.
  \bibinfo{pages}{21271--21284}.
\newblock


\bibitem[He and McAuley(2016a)]%
        {he2016ups}
\bibfield{author}{\bibinfo{person}{Ruining He} {and} \bibinfo{person}{Julian
  McAuley}.} \bibinfo{year}{2016}\natexlab{a}.
\newblock \showarticletitle{Ups and downs: Modeling the visual evolution of
  fashion trends with one-class collaborative filtering}. In
  \bibinfo{booktitle}{\emph{proceedings of the 25th international conference on
  world wide web}}. \bibinfo{pages}{507--517}.
\newblock


\bibitem[He and McAuley(2016b)]%
        {he2016vbpr}
\bibfield{author}{\bibinfo{person}{Ruining He} {and} \bibinfo{person}{Julian
  McAuley}.} \bibinfo{year}{2016}\natexlab{b}.
\newblock \showarticletitle{VBPR: visual bayesian personalized ranking from
  implicit feedback}. In \bibinfo{booktitle}{\emph{Proceedings of the AAAI
  conference on artificial intelligence}}, Vol.~\bibinfo{volume}{30}.
\newblock


\bibitem[He et~al\mbox{.}(2020)]%
        {he2020lightgcn}
\bibfield{author}{\bibinfo{person}{Xiangnan He}, \bibinfo{person}{Kuan Deng},
  \bibinfo{person}{Xiang Wang}, \bibinfo{person}{Yan Li},
  \bibinfo{person}{Yongdong Zhang}, {and} \bibinfo{person}{Meng Wang}.}
  \bibinfo{year}{2020}\natexlab{}.
\newblock \showarticletitle{Lightgcn: Simplifying and powering graph
  convolution network for recommendation}. In
  \bibinfo{booktitle}{\emph{Proceedings of the 43rd International ACM SIGIR
  Conference on Research and Development in Information Retrieval}}.
  \bibinfo{pages}{639--648}.
\newblock


\bibitem[Jaszczur et~al\mbox{.}(2021)]%
        {jaszczur2021sparse}
\bibfield{author}{\bibinfo{person}{Sebastian Jaszczur},
  \bibinfo{person}{Aakanksha Chowdhery}, \bibinfo{person}{Afroz Mohiuddin},
  \bibinfo{person}{Lukasz Kaiser}, \bibinfo{person}{Wojciech Gajewski},
  \bibinfo{person}{Henryk Michalewski}, {and} \bibinfo{person}{Jonni Kanerva}.}
  \bibinfo{year}{2021}\natexlab{}.
\newblock \showarticletitle{Sparse is enough in scaling transformers}.
\newblock \bibinfo{journal}{\emph{Advances in Neural Information Processing
  Systems}}  \bibinfo{volume}{34} (\bibinfo{year}{2021}),
  \bibinfo{pages}{9895--9907}.
\newblock


\bibitem[Jing and Tian(2020)]%
        {jing2020self}
\bibfield{author}{\bibinfo{person}{Longlong Jing} {and} \bibinfo{person}{Yingli
  Tian}.} \bibinfo{year}{2020}\natexlab{}.
\newblock \showarticletitle{Self-supervised visual feature learning with deep
  neural networks: A survey}.
\newblock \bibinfo{journal}{\emph{IEEE transactions on pattern analysis and
  machine intelligence}} \bibinfo{volume}{43}, \bibinfo{number}{11}
  (\bibinfo{year}{2020}), \bibinfo{pages}{4037--4058}.
\newblock


\bibitem[Kingma and Ba(2015)]%
        {kingma2015adam}
\bibfield{author}{\bibinfo{person}{Diederik~P Kingma} {and}
  \bibinfo{person}{Jimmy Ba}.} \bibinfo{year}{2015}\natexlab{}.
\newblock \showarticletitle{Adam: A method for stochastic optimization}. In
  \bibinfo{booktitle}{\emph{International Conference on Learning
  Representations}}.
\newblock


\bibitem[Kipf and Welling(2017)]%
        {kipf2017semi}
\bibfield{author}{\bibinfo{person}{Thomas~N Kipf} {and} \bibinfo{person}{Max
  Welling}.} \bibinfo{year}{2017}\natexlab{}.
\newblock \showarticletitle{Semi-supervised classification with graph
  convolutional networks}. In \bibinfo{booktitle}{\emph{International
  Conference on Learning Representations}}.
\newblock


\bibitem[Lee et~al\mbox{.}(2021)]%
        {lee2021bootstrapping}
\bibfield{author}{\bibinfo{person}{Dongha Lee}, \bibinfo{person}{SeongKu Kang},
  \bibinfo{person}{Hyunjun Ju}, \bibinfo{person}{Chanyoung Park}, {and}
  \bibinfo{person}{Hwanjo Yu}.} \bibinfo{year}{2021}\natexlab{}.
\newblock \showarticletitle{Bootstrapping User and Item Representations for
  One-Class Collaborative Filtering}. In \bibinfo{booktitle}{\emph{Proceedings
  of the 44th International ACM SIGIR Conference on Research and Development in
  Information Retrieval}}.
\newblock


\bibitem[Li et~al\mbox{.}(2021)]%
        {li2021training}
\bibfield{author}{\bibinfo{person}{Guohao Li}, \bibinfo{person}{Matthias
  M{\"u}ller}, \bibinfo{person}{Bernard Ghanem}, {and} \bibinfo{person}{Vladlen
  Koltun}.} \bibinfo{year}{2021}\natexlab{}.
\newblock \showarticletitle{Training graph neural networks with 1000 layers}.
  In \bibinfo{booktitle}{\emph{International conference on machine learning}}.
  PMLR, \bibinfo{pages}{6437--6449}.
\newblock


\bibitem[Liu et~al\mbox{.}(2019)]%
        {liu2019user}
\bibfield{author}{\bibinfo{person}{Fan Liu}, \bibinfo{person}{Zhiyong Cheng},
  \bibinfo{person}{Changchang Sun}, \bibinfo{person}{Yinglong Wang},
  \bibinfo{person}{Liqiang Nie}, {and} \bibinfo{person}{Mohan Kankanhalli}.}
  \bibinfo{year}{2019}\natexlab{}.
\newblock \showarticletitle{User diverse preference modeling by multimodal
  attentive metric learning}. In \bibinfo{booktitle}{\emph{Proceedings of the
  27th ACM international conference on multimedia}}.
  \bibinfo{pages}{1526--1534}.
\newblock


\bibitem[Liu et~al\mbox{.}(2020)]%
        {liu2020towards}
\bibfield{author}{\bibinfo{person}{Meng Liu}, \bibinfo{person}{Hongyang Gao},
  {and} \bibinfo{person}{Shuiwang Ji}.} \bibinfo{year}{2020}\natexlab{}.
\newblock \showarticletitle{Towards deeper graph neural networks}. In
  \bibinfo{booktitle}{\emph{Proceedings of the 26th ACM SIGKDD international
  conference on knowledge discovery \& data mining}}.
  \bibinfo{pages}{338--348}.
\newblock


\bibitem[Liu et~al\mbox{.}(2017)]%
        {liu2017deepstyle}
\bibfield{author}{\bibinfo{person}{Qiang Liu}, \bibinfo{person}{Shu Wu}, {and}
  \bibinfo{person}{Liang Wang}.} \bibinfo{year}{2017}\natexlab{}.
\newblock \showarticletitle{Deepstyle: Learning user preferences for visual
  recommendation}. In \bibinfo{booktitle}{\emph{Proceedings of the 40th
  international acm sigir conference on research and development in information
  retrieval}}. \bibinfo{pages}{841--844}.
\newblock


\bibitem[Liu et~al\mbox{.}(2021)]%
        {liu2021self}
\bibfield{author}{\bibinfo{person}{Xiao Liu}, \bibinfo{person}{Fanjin Zhang},
  \bibinfo{person}{Zhenyu Hou}, \bibinfo{person}{Li Mian},
  \bibinfo{person}{Zhaoyu Wang}, \bibinfo{person}{Jing Zhang}, {and}
  \bibinfo{person}{Jie Tang}.} \bibinfo{year}{2021}\natexlab{}.
\newblock \showarticletitle{Self-supervised learning: Generative or
  contrastive}.
\newblock \bibinfo{journal}{\emph{IEEE Transactions on Knowledge and Data
  Engineering}} (\bibinfo{year}{2021}).
\newblock


\bibitem[Ni et~al\mbox{.}(2019)]%
        {ni2019justifying}
\bibfield{author}{\bibinfo{person}{Jianmo Ni}, \bibinfo{person}{Jiacheng Li},
  {and} \bibinfo{person}{Julian McAuley}.} \bibinfo{year}{2019}\natexlab{}.
\newblock \showarticletitle{Justifying recommendations using distantly-labeled
  reviews and fine-grained aspects}. In \bibinfo{booktitle}{\emph{Proceedings
  of the 2019 Conference on Empirical Methods in Natural Language Processing
  and the 9th International Joint Conference on Natural Language Processing
  (EMNLP-IJCNLP)}}. \bibinfo{pages}{188--197}.
\newblock


\bibitem[Paszke et~al\mbox{.}(2019)]%
        {paszke2019pytorch}
\bibfield{author}{\bibinfo{person}{Adam Paszke}, \bibinfo{person}{Sam Gross},
  \bibinfo{person}{Francisco Massa}, \bibinfo{person}{Adam Lerer},
  \bibinfo{person}{James Bradbury}, \bibinfo{person}{Gregory Chanan},
  \bibinfo{person}{Trevor Killeen}, \bibinfo{person}{Zeming Lin},
  \bibinfo{person}{Natalia Gimelshein}, \bibinfo{person}{Luca Antiga},
  {et~al\mbox{.}}} \bibinfo{year}{2019}\natexlab{}.
\newblock \showarticletitle{Pytorch: An imperative style, high-performance deep
  learning library}.
\newblock \bibinfo{journal}{\emph{Advances in neural information processing
  systems}}  \bibinfo{volume}{32} (\bibinfo{year}{2019}).
\newblock


\bibitem[Reimers and Gurevych(2019)]%
        {reimers2019sentence}
\bibfield{author}{\bibinfo{person}{Nils Reimers} {and} \bibinfo{person}{Iryna
  Gurevych}.} \bibinfo{year}{2019}\natexlab{}.
\newblock \showarticletitle{Sentence-bert: Sentence embeddings using siamese
  bert-networks}. In \bibinfo{booktitle}{\emph{EMNLP}}.
  \bibinfo{pages}{3980--3990}.
\newblock


\bibitem[Rendle et~al\mbox{.}(2009)]%
        {rendle2009bpr}
\bibfield{author}{\bibinfo{person}{Steffen Rendle}, \bibinfo{person}{Christoph
  Freudenthaler}, \bibinfo{person}{Zeno Gantner}, {and} \bibinfo{person}{Lars
  Schmidt-Thieme}.} \bibinfo{year}{2009}\natexlab{}.
\newblock \showarticletitle{BPR: Bayesian Personalized Ranking from Implicit
  Feedback}. In \bibinfo{booktitle}{\emph{Proceedings of the Twenty-Fifth
  Conference on Uncertainty in Artificial Intelligence}}.
  \bibinfo{pages}{452--461}.
\newblock


\bibitem[Shi et~al\mbox{.}(2022)]%
        {shi2022visual}
\bibfield{author}{\bibinfo{person}{Baifeng Shi}, \bibinfo{person}{Yale Song},
  \bibinfo{person}{Neel Joshi}, \bibinfo{person}{Trevor Darrell}, {and}
  \bibinfo{person}{Xin Wang}.} \bibinfo{year}{2022}\natexlab{}.
\newblock \showarticletitle{Visual Attention Emerges from Recurrent Sparse
  Reconstruction}.
\newblock \bibinfo{journal}{\emph{arXiv preprint arXiv:2204.10962}}
  (\bibinfo{year}{2022}).
\newblock


\bibitem[Shorten and Khoshgoftaar(2019)]%
        {shorten2019survey}
\bibfield{author}{\bibinfo{person}{Connor Shorten} {and}
  \bibinfo{person}{Taghi~M Khoshgoftaar}.} \bibinfo{year}{2019}\natexlab{}.
\newblock \showarticletitle{A survey on image data augmentation for deep
  learning}.
\newblock \bibinfo{journal}{\emph{Journal of Big Data}} \bibinfo{volume}{6},
  \bibinfo{number}{1} (\bibinfo{year}{2019}), \bibinfo{pages}{1--48}.
\newblock


\bibitem[Simonyan and Zisserman(2014)]%
        {simonyan2014very}
\bibfield{author}{\bibinfo{person}{Karen Simonyan} {and}
  \bibinfo{person}{Andrew Zisserman}.} \bibinfo{year}{2014}\natexlab{}.
\newblock \showarticletitle{Very deep convolutional networks for large-scale
  image recognition}.
\newblock \bibinfo{journal}{\emph{arXiv preprint arXiv:1409.1556}}
  (\bibinfo{year}{2014}).
\newblock


\bibitem[Srivastava et~al\mbox{.}(2014)]%
        {srivastava2014dropout}
\bibfield{author}{\bibinfo{person}{Nitish Srivastava},
  \bibinfo{person}{Geoffrey Hinton}, \bibinfo{person}{Alex Krizhevsky},
  \bibinfo{person}{Ilya Sutskever}, {and} \bibinfo{person}{Ruslan
  Salakhutdinov}.} \bibinfo{year}{2014}\natexlab{}.
\newblock \showarticletitle{Dropout: a simple way to prevent neural networks
  from overfitting}.
\newblock \bibinfo{journal}{\emph{The journal of machine learning research}}
  \bibinfo{volume}{15}, \bibinfo{number}{1} (\bibinfo{year}{2014}),
  \bibinfo{pages}{1929--1958}.
\newblock


\bibitem[Tarvainen and Valpola(2017)]%
        {tarvainen2017mean}
\bibfield{author}{\bibinfo{person}{Antti Tarvainen} {and}
  \bibinfo{person}{Harri Valpola}.} \bibinfo{year}{2017}\natexlab{}.
\newblock \showarticletitle{Mean teachers are better role models:
  Weight-averaged consistency targets improve semi-supervised deep learning
  results}.
\newblock \bibinfo{journal}{\emph{Advances in neural information processing
  systems}}  \bibinfo{volume}{30}.
\newblock


\bibitem[Thakoor et~al\mbox{.}(2021)]%
        {thakoor2021large}
\bibfield{author}{\bibinfo{person}{Shantanu Thakoor}, \bibinfo{person}{Corentin
  Tallec}, \bibinfo{person}{Mohammad~Gheshlaghi Azar}, \bibinfo{person}{Mehdi
  Azabou}, \bibinfo{person}{Eva~L Dyer}, \bibinfo{person}{Remi Munos},
  \bibinfo{person}{Petar Veli{\v{c}}kovi{\'c}}, {and} \bibinfo{person}{Michal
  Valko}.} \bibinfo{year}{2021}\natexlab{}.
\newblock \showarticletitle{Large-Scale Representation Learning on Graphs via
  Bootstrapping}. In \bibinfo{booktitle}{\emph{International Conference on
  Learning Representations}}.
\newblock


\bibitem[Wang et~al\mbox{.}(2021b)]%
        {wang2021dualgnn}
\bibfield{author}{\bibinfo{person}{Qifan Wang}, \bibinfo{person}{Yinwei Wei},
  \bibinfo{person}{Jianhua Yin}, \bibinfo{person}{Jianlong Wu},
  \bibinfo{person}{Xuemeng Song}, {and} \bibinfo{person}{Liqiang Nie}.}
  \bibinfo{year}{2021}\natexlab{b}.
\newblock \showarticletitle{DualGNN: Dual Graph Neural Network for Multimedia
  Recommendation}.
\newblock \bibinfo{journal}{\emph{IEEE Transactions on Multimedia}}
  (\bibinfo{year}{2021}).
\newblock


\bibitem[Wang et~al\mbox{.}(2021a)]%
        {wang2021graph}
\bibfield{author}{\bibinfo{person}{Shoujin Wang}, \bibinfo{person}{Liang Hu},
  \bibinfo{person}{Yan Wang}, \bibinfo{person}{Xiangnan He},
  \bibinfo{person}{Quan~Z Sheng}, \bibinfo{person}{Mehmet~A Orgun},
  \bibinfo{person}{Longbing Cao}, \bibinfo{person}{Francesco Ricci}, {and}
  \bibinfo{person}{Philip~S Yu}.} \bibinfo{year}{2021}\natexlab{a}.
\newblock \showarticletitle{Graph learning based recommender systems: A
  review}. In \bibinfo{booktitle}{\emph{Proceedings of the 30th International
  Joint Conference on Artificial Intelligence}}. \bibinfo{pages}{4644--4652}.
\newblock


\bibitem[Wei et~al\mbox{.}(2020)]%
        {wei2020graph}
\bibfield{author}{\bibinfo{person}{Yinwei Wei}, \bibinfo{person}{Xiang Wang},
  \bibinfo{person}{Liqiang Nie}, \bibinfo{person}{Xiangnan He}, {and}
  \bibinfo{person}{Tat-Seng Chua}.} \bibinfo{year}{2020}\natexlab{}.
\newblock \showarticletitle{Graph-refined convolutional network for multimedia
  recommendation with implicit feedback}. In
  \bibinfo{booktitle}{\emph{Proceedings of the 28th ACM international
  conference on multimedia}}. \bibinfo{pages}{3541--3549}.
\newblock


\bibitem[Wei et~al\mbox{.}(2019)]%
        {wei2019mmgcn}
\bibfield{author}{\bibinfo{person}{Yinwei Wei}, \bibinfo{person}{Xiang Wang},
  \bibinfo{person}{Liqiang Nie}, \bibinfo{person}{Xiangnan He},
  \bibinfo{person}{Richang Hong}, {and} \bibinfo{person}{Tat-Seng Chua}.}
  \bibinfo{year}{2019}\natexlab{}.
\newblock \showarticletitle{MMGCN: Multi-modal graph convolution network for
  personalized recommendation of micro-video}. In
  \bibinfo{booktitle}{\emph{Proceedings of the 27th ACM International
  Conference on Multimedia}}. \bibinfo{pages}{1437--1445}.
\newblock


\bibitem[Wu et~al\mbox{.}(2021a)]%
        {wu2021self}
\bibfield{author}{\bibinfo{person}{Jiancan Wu}, \bibinfo{person}{Xiang Wang},
  \bibinfo{person}{Fuli Feng}, \bibinfo{person}{Xiangnan He},
  \bibinfo{person}{Liang Chen}, \bibinfo{person}{Jianxun Lian}, {and}
  \bibinfo{person}{Xing Xie}.} \bibinfo{year}{2021}\natexlab{a}.
\newblock \showarticletitle{Self-supervised graph learning for recommendation}.
  In \bibinfo{booktitle}{\emph{Proceedings of the 44th international ACM SIGIR
  conference on research and development in information retrieval}}.
  \bibinfo{pages}{726--735}.
\newblock


\bibitem[Wu et~al\mbox{.}(2021b)]%
        {wu2020self}
\bibfield{author}{\bibinfo{person}{Jiancan Wu}, \bibinfo{person}{Xiang Wang},
  \bibinfo{person}{Fuli Feng}, \bibinfo{person}{Xiangnan He},
  \bibinfo{person}{Liang Chen}, \bibinfo{person}{Jianxun Lian}, {and}
  \bibinfo{person}{Xing Xie}.} \bibinfo{year}{2021}\natexlab{b}.
\newblock \showarticletitle{Self-supervised Graph Learning for Recommendation}.
  In \bibinfo{booktitle}{\emph{Proceedings of the 44rd International ACM SIGIR
  Conference on Research and Development in Information Retrieval}}.
\newblock


\bibitem[Wu et~al\mbox{.}(2020)]%
        {wu2020graph}
\bibfield{author}{\bibinfo{person}{Shiwen Wu}, \bibinfo{person}{Fei Sun},
  \bibinfo{person}{Wentao Zhang}, \bibinfo{person}{Xu Xie}, {and}
  \bibinfo{person}{Bin Cui}.} \bibinfo{year}{2020}\natexlab{}.
\newblock \showarticletitle{Graph neural networks in recommender systems: a
  survey}.
\newblock \bibinfo{journal}{\emph{ACM Computing Surveys (CSUR)}}
  (\bibinfo{year}{2020}).
\newblock


\bibitem[Zbontar et~al\mbox{.}(2021)]%
        {zbontar2021barlow}
\bibfield{author}{\bibinfo{person}{Jure Zbontar}, \bibinfo{person}{Li Jing},
  \bibinfo{person}{Ishan Misra}, \bibinfo{person}{Yann LeCun}, {and}
  \bibinfo{person}{St{\'e}phane Deny}.} \bibinfo{year}{2021}\natexlab{}.
\newblock \showarticletitle{Barlow twins: Self-supervised learning via
  redundancy reduction}.
\newblock \bibinfo{journal}{\emph{arXiv preprint arXiv:2103.03230}}
  (\bibinfo{year}{2021}).
\newblock


\bibitem[Zhang et~al\mbox{.}(2021)]%
        {zhang2021mining}
\bibfield{author}{\bibinfo{person}{Jinghao Zhang}, \bibinfo{person}{Yanqiao
  Zhu}, \bibinfo{person}{Qiang Liu}, \bibinfo{person}{Shu Wu},
  \bibinfo{person}{Shuhui Wang}, {and} \bibinfo{person}{Liang Wang}.}
  \bibinfo{year}{2021}\natexlab{}.
\newblock \showarticletitle{Mining Latent Structures for Multimedia
  Recommendation}. In \bibinfo{booktitle}{\emph{Proceedings of the 29th ACM
  International Conference on Multimedia}}. \bibinfo{pages}{3872--3880}.
\newblock


\bibitem[Zhang et~al\mbox{.}(2022)]%
        {zhang2022diffusion}
\bibfield{author}{\bibinfo{person}{Lingzi Zhang}, \bibinfo{person}{Yong Liu},
  \bibinfo{person}{Xin Zhou}, \bibinfo{person}{Chunyan Miao},
  \bibinfo{person}{Guoxin Wang}, {and} \bibinfo{person}{Haihong Tang}.}
  \bibinfo{year}{2022}\natexlab{}.
\newblock \showarticletitle{Diffusion-based graph contrastive learning for
  recommendation with implicit feedback}. In \bibinfo{booktitle}{\emph{Database
  Systems for Advanced Applications: 27th International Conference, DASFAA
  2022, Virtual Event, April 11--14, 2022, Proceedings, Part II}}. Springer,
  \bibinfo{pages}{232--247}.
\newblock


\bibitem[Zhang et~al\mbox{.}(2019)]%
        {zhang2019deep}
\bibfield{author}{\bibinfo{person}{Shuai Zhang}, \bibinfo{person}{Lina Yao},
  \bibinfo{person}{Aixin Sun}, {and} \bibinfo{person}{Yi Tay}.}
  \bibinfo{year}{2019}\natexlab{}.
\newblock \showarticletitle{Deep learning based recommender system: A survey
  and new perspectives}.
\newblock \bibinfo{journal}{\emph{ACM Computing Surveys (CSUR)}}
  \bibinfo{volume}{52}, \bibinfo{number}{1} (\bibinfo{year}{2019}),
  \bibinfo{pages}{1--38}.
\newblock


\bibitem[Zhang et~al\mbox{.}(2013)]%
        {zhang2013optimizing}
\bibfield{author}{\bibinfo{person}{Weinan Zhang}, \bibinfo{person}{Tianqi
  Chen}, \bibinfo{person}{Jun Wang}, {and} \bibinfo{person}{Yong Yu}.}
  \bibinfo{year}{2013}\natexlab{}.
\newblock \showarticletitle{Optimizing top-n collaborative filtering via
  dynamic negative item sampling}. In \bibinfo{booktitle}{\emph{Proceedings of
  the 36th international ACM SIGIR conference on Research and development in
  information retrieval}}. \bibinfo{pages}{785--788}.
\newblock


\bibitem[Zhou et~al\mbox{.}(2023a)]%
        {zhou2023comprehensive}
\bibfield{author}{\bibinfo{person}{Hongyu Zhou}, \bibinfo{person}{Xin Zhou},
  \bibinfo{person}{Zhiwei Zeng}, \bibinfo{person}{Lingzi Zhang}, {and}
  \bibinfo{person}{Zhiqi Shen}.} \bibinfo{year}{2023}\natexlab{a}.
\newblock \showarticletitle{A Comprehensive Survey on Multimodal Recommender
  Systems: Taxonomy, Evaluation, and Future Directions}.
\newblock \bibinfo{journal}{\emph{arXiv preprint arXiv:2302.04473}}
  (\bibinfo{year}{2023}).
\newblock


\bibitem[Zhou et~al\mbox{.}(2023b)]%
        {zhou2023enhancing}
\bibfield{author}{\bibinfo{person}{Hongyu Zhou}, \bibinfo{person}{Xin Zhou},
  \bibinfo{person}{Lingzi Zhang}, {and} \bibinfo{person}{Zhiqi Shen}.}
  \bibinfo{year}{2023}\natexlab{b}.
\newblock \showarticletitle{Enhancing Dyadic Relations with Homogeneous Graphs
  for Multimodal Recommendation}.
\newblock \bibinfo{journal}{\emph{arXiv preprint arXiv:2301.12097}}
  (\bibinfo{year}{2023}).
\newblock


\bibitem[Zhou(2022)]%
        {zhou2022tale}
\bibfield{author}{\bibinfo{person}{Xin Zhou}.} \bibinfo{year}{2022}\natexlab{}.
\newblock \showarticletitle{A Tale of Two Graphs: Freezing and Denoising Graph
  Structures for Multimodal Recommendation}.
\newblock \bibinfo{journal}{\emph{arXiv preprint arXiv:2211.06924}}
  (\bibinfo{year}{2022}).
\newblock


\bibitem[Zhou(2023)]%
        {zhou2023mmrecsm}
\bibfield{author}{\bibinfo{person}{Xin Zhou}.} \bibinfo{year}{2023}\natexlab{}.
\newblock \showarticletitle{MMRec: Simplifying Multimodal Recommendation}.
\newblock \bibinfo{journal}{\emph{arXiv preprint arXiv:2302.03497}}
  (\bibinfo{year}{2023}).
\newblock


\bibitem[Zhou et~al\mbox{.}(2022)]%
        {zhou2022layer}
\bibfield{author}{\bibinfo{person}{Xin Zhou}, \bibinfo{person}{Donghui Lin},
  \bibinfo{person}{Yong Liu}, {and} \bibinfo{person}{Chunyan Miao}.}
  \bibinfo{year}{2022}\natexlab{}.
\newblock \showarticletitle{Layer-refined Graph Convolutional Networks for
  Recommendation}.
\newblock \bibinfo{journal}{\emph{arXiv preprint arXiv:2207.11088}}
  (\bibinfo{year}{2022}).
\newblock


\bibitem[Zhou et~al\mbox{.}(2021)]%
        {zhou2021selfcf}
\bibfield{author}{\bibinfo{person}{Xin Zhou}, \bibinfo{person}{Aixin Sun},
  \bibinfo{person}{Yong Liu}, \bibinfo{person}{Jie Zhang}, {and}
  \bibinfo{person}{Chunyan Miao}.} \bibinfo{year}{2021}\natexlab{}.
\newblock \showarticletitle{SelfCF: A Simple Framework for Self-supervised
  Collaborative Filtering}.
\newblock \bibinfo{journal}{\emph{arXiv preprint arXiv:2107.03019}}
  (\bibinfo{year}{2021}).
\newblock


\end{thebibliography}

\end{document}